\documentclass[lettersize,journal]{IEEEtran}
\usepackage{amsmath,amsfonts}
\usepackage{algorithmic}
\usepackage{algorithm}
\usepackage{array}
\usepackage[caption=false,font=footnotesize,labelfont=sf,textfont=sf]{subfig}
\usepackage{textcomp}
\usepackage{stfloats}
\usepackage{url}
\usepackage{verbatim}
\usepackage{graphicx}
\usepackage{cite}
\usepackage[hidelinks,colorlinks=true,linkcolor=blue,citecolor=blue,urlcolor=black]{hyperref}
\usepackage{orcidlink}

\usepackage{xcolor}
\usepackage{pifont} 
\usepackage{multicol, multirow}
\usepackage{color,soul}
\usepackage{enumitem}
\usepackage{makecell, colortbl}

\usepackage{tikz}
\usepackage{collcell}
\usepackage{booktabs}
\usepackage{siunitx}

\usepackage{tabularx}
\usepackage{float}

\usepackage{amsthm}
\usepackage{mathtools}  

\usepackage{listings, minted}

\usepackage{tcolorbox}
\usepackage{xspace, xstring}
\usepackage{fontawesome}
\usepackage{wrapfig}
\usepackage[percent]{overpic}

\begin{document}

\colorlet{revisedcolor}{black}
\colorlet{xiaoningrv}{violet}

\newcommand{\ehl}[1]{\textcolor{revisedcolor}{#1}}
\newcommand{\truevul}{\textsc{UntrustVul}}
\newcommand{\textcode}[1]{\texttt{#1}}

\newtheorem{dfn}{Definition}

\setul{1pt}{.4pt}

\newcommand{\editnote}[3]{}

\newcommand{\lam}[1]{\editnote{purple}{Lam}{#1}}
\newcommand{\aldeida}[1]{\editnote{red}{Aldeida}{#1}}
\newcommand{\xiaoning}[1]{\editnote{brown}{Xiaoning}{#1}}
\newcommand{\neelofar}[1]{\editnote{teal}{Neelofar}{#1}}
    
\NewTColorBox{custombox}{o}{
    colback=cyan!5,
	colframe=cyan,
	left=1.0mm,
	right=1.0mm,
	top=1mm,
	bottom=1mm,
	fonttitle=\bfseries,
	arc=0mm,
	leftrule=0mm,
	rightrule=0mm,
	toprule=0mm,
	bottomrule=0mm,
    boxrule=0.7pt,
	notitle,
	before=\par\medskip\noindent,
    IfValueTF={#1}{before upper={\faHandORight{}~\textbf{#1: } }}{},
    parbox=false,
}

\newcommand*{\DanNumber}{0.4}%
\newcommand*{\MinNumber}{0.55}%
\newcommand*{\MidNumber}{0.65}%
\newcommand*{\MaxNumber}{1.0}%

\newcommand{\ApplyGradient}[1]{%
        \IfDecimal{#1}{
            \ifdim #1 pt > \MidNumber pt
                \pgfmathparse{(#1-\MidNumber)/(\MaxNumber-\MidNumber+0.001)*100}%
                \global\let\colorratio\pgfmathresult
                \cellcolor{blue!\colorratio}%
                \ifdim #1 pt > 0.85 pt
                    \textcolor{white}{#1}
                \else
                    \textcolor{black}{#1}
                \fi
            \else
                \ifdim #1 pt > \MinNumber pt
                    \pgfmathparse{(#1-\MinNumber)/(\MaxNumber*2)*100}%
                    \global\let\colorratio\pgfmathresult
                    \cellcolor{blue!\colorratio}%
                    \textcolor{black}{#1}
                \else
                    \ifdim #1 pt = 0 pt
                        \cellcolor{red}%
                        \textcolor{black}{#1}
                    \else
                        \ifdim #1 pt < \DanNumber pt
                            \pgfmathparse{(\DanNumber-#1)/(\DanNumber)*100}%
                            \global\let\colorratio\pgfmathresult
                            \cellcolor{red!\colorratio}%
                            \textcolor{black}{#1}
                        \else
                            \cellcolor{white}%
                            \textcolor{black}{#1}
                        \fi
                    \fi
                \fi
            \fi
        }{
            \cellcolor{white}%
            \textcolor{black}{#1}
        }
}

\newcommand{\ApplyGradientReverse}[1]{%
        \IfDecimal{#1}{
            \pgfmathparse{#1*100}%
            \global\let\colorratio\pgfmathresult
            \cellcolor{red!\colorratio}%
            \textcolor{black}{#1}
        }{
            \cellcolor{gray!15}%
            \textcolor{black}{}
        }
}

\ExplSyntaxOn

\NewExpandableDocumentCommand{\gobblefirst}{m}
 {
  \tl_tail:n { #1 }
 }

\ExplSyntaxOff

\newcommand{\ApplyGradientDiff}[1]{%
        \IfDecimal{#1}{
            \ifdim #1 pt < 0 pt
                \cellcolor{red!30}%
                \textcolor{black}{+\gobblefirst{#1}}
            \else
                \pgfmathparse{#1*120pt}%
                \global\let\colorratio\pgfmathresult
                \cellcolor{blue!\colorratio}%
                \textcolor{black}{-\gobblefirst{#1}}
            \fi
        }{
            \cellcolor{white}%
            \textcolor{black}{#1}
        }
}

\newcolumntype{C}{>{\collectcell\ApplyGradient}c<{\endcollectcell}}
\newcolumntype{P}{>{\collectcell\ApplyGradientReverse}c<{\endcollectcell}}
\newcolumntype{D}{@{}>{\collectcell\ApplyGradientDiff}c<{\endcollectcell}}

\title{\truevul: Automated Untrustworthy Alert Identification in Vulnerability Detection Models}

\author{Lam Nguyen Tung \orcidlink{0009-0000-3038-8403}, \textit{Graduate Student Member, IEEE,} 
    Xiaoning Du \orcidlink{0000-0003-3728-9541}, \textit{Member, IEEE},
    Neelofar Neelofar \orcidlink{0000-0003-2572-0250}, 
    Aldeida Aleti \orcidlink{0000-0002-1716-690X}, \textit{Associate Member, IEEE}
}



\maketitle

\begin{abstract}
Machine learning (ML) has shown promising results in detecting software vulnerabilities. 
However, ML detectors are not guaranteed to make predictions based on the right indicators.
Studies have revealed that they can rely on \textit{irrelevant} code features, such as identifiers or function signatures, particularly those that commonly appear in vulnerable code, yet are not related to the actual vulnerabilities.
As a result, the lines of code that the detectors depend on and flag as suspicious are not always genuinely vulnerable.
Consequently, developers must manually review these suspicious lines,
which is 
time-consuming and error-prone.
If the suspicious lines are wrong, developers may be misled, spend unnecessary effort, or even reach incorrect patching strategies.
This highlights the need for automated approaches to identify untrustworthy vulnerability predictions.

In this paper, we introduce \truevul, a new approach for identifying untrustworthy vulnerability predictions.
Specifically, we focus on cases where a model highlights suspicious lines that would not appear in reliable predictions, i.e., lines that are inherently non-vulnerable and unrelated to any vulnerabilities.
To achieve this, we 
leverage patterns of vulnerable lines observed in historical data.
\truevul{} automatically rules out as untrustworthy any predictions that highlight suspicious lines neither observed in history nor influential to those that have been observed.
We refer to such lines as vulnerability-irrelevant.
A line is deemed vulnerability-irrelevant if 
\ding{172}~it does not match any known patterns of historical vulnerabilities,
and \ding{173}~all its successors in the data and control dependency graph are also vulnerability-irrelevant. 
Intuitively, a vulnerability-irrelevant line shows low similarity to known vulnerabilities and has no dependency paths to any lines outside the vulnerability-irrelevant category. 
Notably, these rules are designed to be conservative, as mislabeling a trustworthy prediction as untrustworthy is also undesired.
We evaluate \truevul{} on 115K vulnerability predictions made by four models across BigVul, MegaVul, SARD, and PrimeVul datasets, with ground-truth trustworthiness labeled based on the overlap between actual denoised vulnerable lines and model-annotated suspicious lines.
\truevul{} effectively detects untrustworthy predictions with AUC of 70\%--88\%
and F1-score of 82\%--94\%, outperforming existing approaches by 6\%--59\% in AUC and 13\%--92\% in F1-score.
\end{abstract}


\section{Introduction}
\label{sec:introduction}

\begin{figure}[ht]
    \centering
    \begin{overpic}[width=\linewidth, trim={0 21.5cm 0 0}, clip]{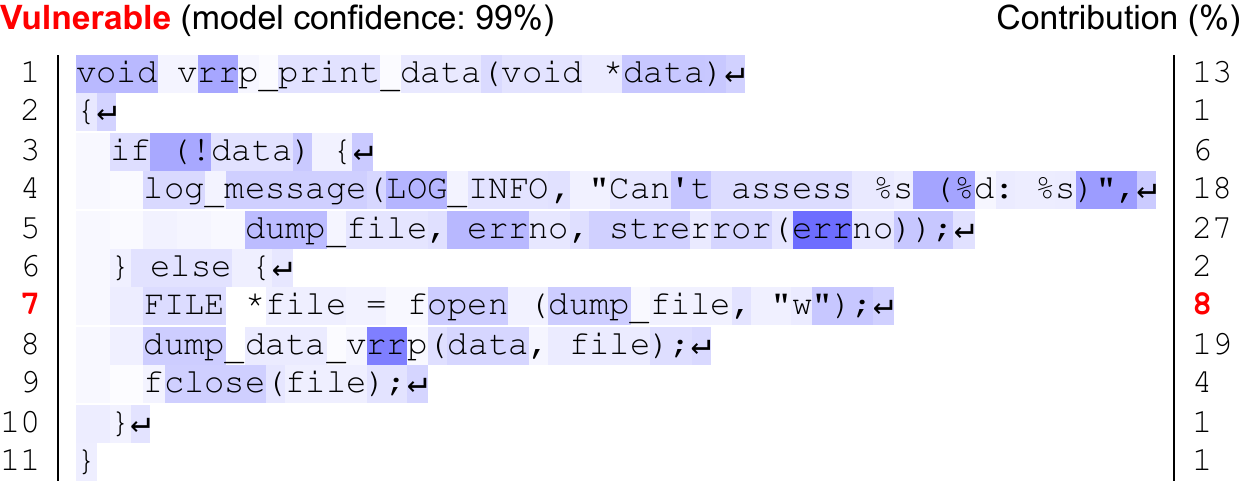}
    \put(55,0){\includegraphics[scale=0.5, trim={0 0 0 0}, clip]{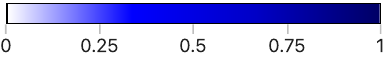}}
    \end{overpic}
    
    
    \caption{\color{revisedcolor} A prediction made by LineVul~\cite{Fu:2022:LineVul:ATransformerBasedLineLevelVulnerabilityPrediction} with its attention-based interpretation at the line level.
    Darker shading indicates tokens with more significant contributions.
    Each line's contribution shown on the right side is the sum of all token-level contributions within that line.
    }
    \label{fig:intro}
\end{figure}

\IEEEPARstart{M}{achine} learning (ML) has been widely used to detect vulnerabilities, i.e.,
weaknesses in software code that can be exploited by attackers~\cite{Johnson:2021:GuideforSecurityFocusedConfigurationManagementofInformationSystems}.
Recent studies~\cite{Zhou:2019:Devign:EffectiveVulnerabilityIdentificationbyLearningComprehensiveProgramSemanticsviaGraphNeuralNetworks, Fu:2022:LineVul:ATransformerBasedLineLevelVulnerabilityPrediction, Chakraborty:2022:DeepLearningBasedVulnerabilityDetectionAreWeThereYet, Cheng:2021:DeepWukong:StaticallyDetectingSoftwareVulnerabilitiesUsingDeepGraphNeuralNetwork} have shown good results, with state-of-the-art (SOTA) models achieving up to 95\% accuracy in held-out evaluations for detecting vulnerable functions.
To review these detected vulnerabilities and identify root causes~\cite{Stradowski:2024:InterpretabilityExplainabilityAppliedtoMachineLearningSoftwareDefectPredictionAnIndustrialPerspective, Steenhoek:2025:ClosingtheGapAUserStudyontheRealworldUsefulnessofAIpoweredVulnerabilityDetectionRepairintheIDE}, developers need to manually assess suspicious lines of code highlighted by fine-grained interpretability techniques~\cite{Li:2021:VulnerabilityDetectionwithFineGrainedInterpretations, Zhang:2023:LearningtoLocateandDescribeVulnerabilities, Cheng:2024:BeyondFidelityExplainingVulnerabilityLocalizationofLearningBasedDetectors}.
However, such interpretations may not accurately explain why a function is vulnerable, thereby failing to pinpoint the root causes.

Investigating their behaviors, studies~\cite{Ganz:2021:ExplainingGraphNeuralNetworksforVulnerabilityDiscovery, Chakraborty:2022:DeepLearningBasedVulnerabilityDetectionAreWeThereYet, Steenhoek:2023:AnEmpiricalStudyofDeepLearningModelsforVulnerabilityDetection, Cheng:2024:BeyondFidelityExplainingVulnerabilityLocalizationofLearningBasedDetectors} have found that ML models often learn spurious correlations during training. 
These correlations are tied to irrelevant code parts, such as 
identifiers, coding styles, or comments that frequently appear in vulnerable code.
As a result, the models may correctly identify vulnerable functions by chance, but the reasons they rely on are incorrect.
In particular, they may rely on patterns that are not semantically relevant to the underlying vulnerabilities.
Hence, their interpretations can give misleading reasons for the vulnerabilities.
This hinders developers from locating the root causes and increases the burden of manual assessment.
Worse, following these interpretations risks creating flawed patches that fail to address the true vulnerabilities, or even introduce new ones.

A vulnerability prediction is untrustworthy if it highlights lines of code (hereafter referred to as lines) that are unrelated to the actual 
vulnerabilities.
For example, the function \textcode{vrrp\_print\_data} in Figure~\ref{fig:intro} contains a CWE-59\footnote{
Improper link resolution before file access (link following) vulnerability,  
\url{https://cwe.mitre.org/data/definitions/59.html}} vulnerability at Line~7, which accesses a file without ensuring it cannot be redirected to an unintended resource.
This function is correctly predicted as vulnerable by LineVul~\cite{Fu:2022:LineVul:ATransformerBasedLineLevelVulnerabilityPrediction}, an SOTA model based on CodeBERT~\cite{Feng:2020:CodeBERT:APreTrainedModelforProgrammingandNaturalLanguages}.
The model confidence 
is measured by the model-calculated probability that the function is vulnerable.
The figure also provides the fine-grained interpretation generated by the attention mechanism.
This is a feature provided by LineVul, with its annotation displayed in the figure.
In particular, tokens' darker shading indicates more significant contributions.
Each line’s contribution shown on the figure's right side is summarised from all tokens in that line.
This allows us to identify suspicious lines that contribute the most to the prediction.
Although LineVul correctly 
predicts this function as vulnerable 
with a confidence of nearly 100\%, the most suspicious lines annotated by this prediction are mostly irrelevant.
Line~5 (27\%) and Line~4 (18\%) are from a different branch than Line~7 (8\%), the actual vulnerable line.
Similarly, Line~8 (19\%) only flows to Line~9, which is non-vulnerable.
The function signature at Line~1 (13\%) contributes even more than the vulnerable line.
We argue this prediction is \textit{untrustworthy}, as it fails to pinpoint vulnerability-related lines.
Developers may follow the suspicious lines, mistakenly associating the variable \textcode{data} with the vulnerability.
They can create patches that fail to address the vulnerability at Line~7 or even introduce new ones.
For this reason, automated approaches are needed to detect untrustworthy predictions.
These approaches prevent overlooked vulnerabilities and alleviate the burden of manual assessment.
However, developing such approaches is challenging due to the complex and dynamic nature of code.
Hence, in this work, we focus on cases of model-annotated suspicious lines of code that should not occur in trustworthy predictions.

\subsubsection*{\textbf{\ul{Methodology}}}
We introduce \truevul, the first automated approach that can expose untrustworthy vulnerability predictions.
It takes as input a vulnerability prediction during inference that annotates suspicious lines.
The prediction's trustworthiness is then assessed based on whether these suspicious lines are \textit{vulnerability-irrelevant}.
A line is vulnerability-irrelevant if it \ding{172}~itself is extremely unlikely to be vulnerable and \ding{173}~has no influence on potentially vulnerable lines.

A key insight is that patterns of vulnerable lines have been extensively documented in historical data.
Based on this, \truevul{} assesses whether a line itself cannot be vulnerable, which we refer to as a \textit{benign candidate}, by observing historical data.
In contrast, its counterpart is called non-benign candidates.
Benign candidates are context-free non-vulnerable, meaning they are not observed in known vulnerable lines.
However, they may still influence vulnerabilities.
Accurately finding all vulnerability-influencing lines is challenging, as ground-truth vulnerabilities are unknown.
To handle this, we treat non-benign candidates as an overestimation of actual vulnerabilities. 
These non-benign candidates are potentially vulnerable, which may or may not be the actual vulnerabilities, depending on their control or data dependencies.
Benign candidates that do not reach any non-benign ones are classified as vulnerability-irrelevant.
Hence, we consider
a line as vulnerability-irrelevant if it \ding{172}~exhibits no patterns of historical vulnerable lines, and \ding{173}~does not reach other suspicious lines that exhibit vulnerability patterns via control or data dependencies.
Notably, these rules are deliberately conservative, as mislabeling a trustworthy prediction as untrustworthy is also undesired.

Assessing \ding{172} can be done by determining whether the line is a benign candidate, not matching any patterns of historical vulnerable lines.
Assessing \ding{173} requires analysing reachability
via 
dependencies that influence control flow or manipulate variables associated with non-benign candidates.
Specifically, \truevul{} identifies benign candidates among the prediction-annotated suspicious lines by comparing their deep representations with those of historical vulnerable lines.
The remaining suspicious lines are non-benign candidates likely to be vulnerable, potentially including actual vulnerabilities. 
This approach can misclassify some vulnerability-irrelevant lines as non-benign candidates.
However, our empirical evaluation in Section~\ref{sec:expr_results} shows that the misclassification rate is low.

We further filter out vulnerability-influencing lines from benign candidates to refine the group of vulnerability-irrelevant lines.
Benign candidates that do not reach any non-benign candidates are naturally deemed as irrelevant, and predictions relying on them are definitely untrustworthy.
To do this, we trace
dependencies of the benign candidates to check whether they can reach
a non-benign candidate
using static and rule-based analyses.
These analyses are inherently precise,
which allows \truevul{} to achieve high Precision in detecting untrustworthy predictions with minimal false positives. 
A caveat is that Recall may be relatively lower, as some edge cases may be missed.
Nevertheless, Section~\ref{sec:expr_results} shows that \truevul{} maintains high Recall and successfully handles a considerable number of edge cases.

To formalise the analyses, we define multiple rules for \textit{relevant dependencies}.
We also introduce the concept of \textit{reachability distance}, which measures the length of a sequence of relevant dependencies from a benign candidate to the nearest non-benign one.
This estimates the strength of the benign candidate's influence on potential vulnerabilities.
Finally, a score is calculated based on the reachability distance of the suspicious lines and their contribution levels to the prediction.
Instead of being binary, this score is continuous, capturing the spectrum of trustworthiness.
A lower score indicates a less trustworthy prediction.


\subsubsection*{\textbf{\ul{Significance}}}
Extensive studies~\cite{Chu:2024:GraphNeuralNetworksforVulnerabilityDetectionACounterfactualExplanation, Fu:2022:LineVul:ATransformerBasedLineLevelVulnerabilityPrediction, Cheng:2024:BeyondFidelityExplainingVulnerabilityLocalizationofLearningBasedDetectors} have investigated how to provide interpretations behind ML vulnerability predictions.
In contrast, this paper is the first to leverage such interpretations for the automated detection of untrustworthy predictions.
With the increasing adoption of ML for vulnerability detection, 
developers face a growing burden of manually reviewing ML predictions. This is aggravated
by untrustworthy predictions, even with high model confidence.
We mitigate this challenge by automatically distinguishing untrustworthy predictions 
from those that warrant closer inspection.
This prevents developers from being misled by untrustworthiness.
Crucially, \truevul{} does not require knowing vulnerability locations. 
Instead, it analyses model-annotated suspicious lines using independent components, making it applicable and effective across different models in real-world settings.

\subsubsection*{\textbf{\ul{Evaluation}}}
We evaluate \truevul{} on four SOTA vulnerability detectors, LineVul~\cite{Fu:2022:LineVul:ATransformerBasedLineLevelVulnerabilityPrediction}, SVulD~\cite{Ni:2023:DistinguishingLookAlikeInnocentandVulnerableCodebySubtleSemanticRepresentationLearningandExplanation}, ReVeal~\cite{Chakraborty:2022:DeepLearningBasedVulnerabilityDetectionAreWeThereYet}, and IVDetect~\cite{Li:2021:VulnerabilityDetectionwithFineGrainedInterpretations},
using BigVul~\cite{Fan:2020:BigVul:ACCCodeVulnerabilityDatasetwithCodeChangesandCVESummaries}, MegaVul~\cite{Ni:2024:MegaVul:AC/CppVulnerabilityDatasetwithComprehensiveCodeRepresentations}, SARD~\cite{SARD}, and PrimeVul~\cite{Ding:2024:VulnerabilityDetectionwithCodeLanguageModelsHowFarAreWe} datasets.
We first collect predictions made by the models on these datasets.
Then, ground-truth trustworthiness labels of predictions are derived from the overlap between actual denoised vulnerable lines and prediction-annotated suspicious lines.
These ground-truth trustworthiness labels are then used to verify \truevul.
Results show that 
\truevul{} achieves F1-scores of 82\%–94\% in detecting untrustworthy predictions.
It outperforms an approach based solely on model confidence by 1\%–62\% and CausalVul~\cite{Rahman:2024:TowardsCausalDeepLearningforVulnerabilityDetection} that perturbs code to alter predictions by 28\%–90\%.
\truevul{} also improves vulnerability detectors by up to 321\% in F1-score and 100\% in trustworthiness.

{\textbf{\ul{\textit{Contributions}}}} in this paper can be summarised as follows.

\begin{itemize}
    \item \truevul{}, the first automated approach that can expose untrustworthy vulnerability predictions.
    \item An investigation of the relation between prediction certainty and trustworthiness, showing that prediction certainty is insufficient to detect untrustworthy predictions.
    \item Comprehensive evaluations of \truevul{}, showcasing its superior effectiveness in detecting untrustworthy predictions and improving vulnerability detectors.
    \item Our code and datasets are publicly available online at \textbf{https://doi.org/10.5281/zenodo.15031367}.
\end{itemize}

\section{Definition and Motivation}
\label{sec:motivation}
This section clarifies the definition of trustworthiness and 
the motivation
for detecting untrustworthy predictions.

\subsection{Glossary}

\ehl{This paper uses numerous technical and scholarly concepts. 
To enhance readability, we define each concept upon its first appearance. We then use the corresponding terminology consistently throughout the paper.}
\ehl{For easy reference, Table~\ref{table:glossary} presents a complete glossary of all terms and their definitions.}
\begin{table}[!ht]
\centering
\vspace{-5mm}
{\color{revisedcolor}
\caption{Glossary}
\label{table:glossary}
\resizebox{\linewidth}{!}{
\begin{tabular}{|>{\centering\arraybackslash}p{1.5cm}|p{6.5cm}|}
\hline
\multicolumn{1}{|c}{\textbf{Term}} & \multicolumn{1}{|c|}{\textbf{Definition}} \\
\hline
\multirow{3}{1.5cm}{\centering Suspicious line} &  A suspicious line is a line of code that the vulnerability detector relies on for prediction, which can be identified by fine-grained line-level interpretability techniques. \\
\hline
\multirow{3}{1.5cm}{\centering Benign candidate} &  {A benign candidate is an individual line of code that is context-free benign, meaning it is almost never observed in historical vulnerable lines.}
\\
\hline
\multirow{3}{1.5cm}{\centering Non-benign candidate} & Non-benign candidates are the counterparts of benign candidates. They are lines containing sensitive operations that can trigger vulnerabilities, as observed in history.\\
\hline
\multirow{3}{1.5cm}{\centering Vulnerability-irrelevant} & {A line of code is vulnerability-irrelevant if it itself is a benign candidate, and has no influence on any non-benign candidate via control or data dependencies.}
\\
\hline
\multirow{3}{1.5cm}{\centering Importance score} & In a line-level interpretation of a vulnerability prediction, the importance score quantifies a suspicious line's contribution to the prediction.                  \\
\hline
\multirow{6}{1.5cm}{\centering Program Dependency Graph (PDG)} & A program dependency graph (PDG) is a directed graph explicitly representing dependencies among statements and predicate expressions. They are constructed using two types of edges: data dependencies reflecting the influence of one variable on another, and control dependencies corresponding to the control flow between statements.                  \\
\hline
\multirow{6}{1.5cm}{\centering Intersection over Union (IoU)} & Intersection over union (IoU)~\cite{Hu:2023:InterpretersforGNNBasedVulnerabilityDetectionAreWeThereYet} is a popular metric for localization accuracy. It measures the overlap between a predicted region and the ground truth (actual) one. 
In this study, the overlap is between suspicious lines ($E$) annotated by the prediction and ground-truth vulnerable lines ($G$), IoU = \(\frac{|E\;\cap\;G|}{|E\;\cup\;G|}\).
\\
\hline
\end{tabular}%
}
}
\end{table}

\subsection{Trustworthiness in Vulnerability Detection}

Trustworthiness is a complex concept, making it challenging for researchers to agree on a unified definition~\cite{Kaur:2022:TrustworthyArtificialIntelligenceAReview, Li:2023:TrustworthyAIFromPrinciplestoPractices}. 
Although it is not easy to choose a specific definition, we follow the definition proposed by Lam et al.~\cite{Lam:2024:AutomatedTrustworthinessOracleGenerationforMachineLearningTextClassifiers}. Specifically, we consider a trustworthy prediction to be (a) correct and (b) the reasoning behind it is also plausible.
We refine this definition for vulnerability detection.
For (a), ML predictions, which classify a function as vulnerable or not, should correctly detect vulnerable functions.
For (b), 
such predictions should annotate lines of code related to
the actual vulnerabilities~\cite{Stradowski:2024:InterpretabilityExplainabilityAppliedtoMachineLearningSoftwareDefectPredictionAnIndustrialPerspective}.

We refer to \textit{vulnerability predictions}, or shortly \textit{predictions}, as classifications that label code as vulnerable and annotate specific lines of code as 
suspicious. 
Accordingly, we define an \textit{untrustworthy vulnerability prediction} as follows.
\begin{dfn} 
\label{def:trustworthy}
An untrustworthy prediction correctly flags a vulnerable function but highlights irrelevant lines of code.
\end{dfn}
\noindent
We consider a line irrelevant if it neither exhibits patterns of vulnerable lines nor reaches lines that do exhibit via control or data dependencies.
This approach may not exhaustively capture all untrustworthy predictions.
However, it can precisely detect them with minimal false positives.
\begin{figure}[t]
    \centering
    {\color{revisedcolor}
    \begin{overpic}[width=\linewidth, trim={0 21.5cm 0 0}, clip]{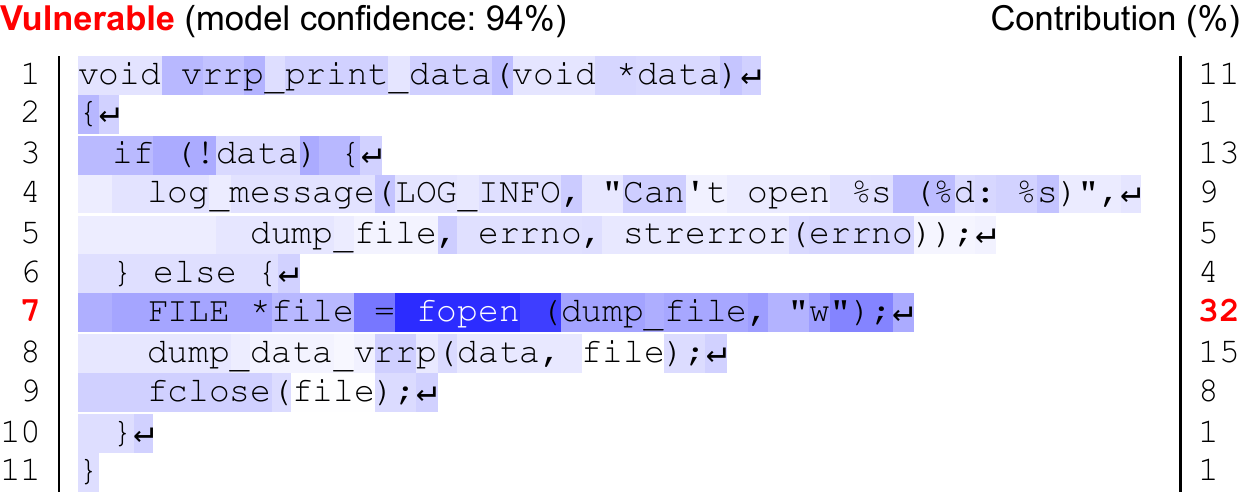}
    \put(55,0){\includegraphics[scale=0.5, trim={0 0 0 0}, clip]{figures/heatmap_scale.pdf}}
    \end{overpic}

    \caption{A trustworthy prediction made by SVulD~\cite{Ni:2023:DistinguishingLookAlikeInnocentandVulnerableCodebySubtleSemanticRepresentationLearningandExplanation} with its attention-based interpretation at the line level.
    Darker shading indicates tokens with more significant contributions.
    Each line's contribution shown on the right side is the sum of all token-level contributions within that line.
    }
    \label{fig:trustworthypred}
    }
\end{figure}
\ehl{For example, the prediction of LineVul~\cite{Fu:2022:LineVul:ATransformerBasedLineLevelVulnerabilityPrediction} shown in Figure~\ref{fig:intro} mainly highlights Lines~4,~5, and~8, particularly tokens \textcode{err} and \textcode{rr}. 
As discussed in Section~\ref{sec:introduction}, they are irrelevant to the vulnerability at Line~7, making the prediction untrustworthy.
In contrast, to illustrate a trustworthy counterpart, we present another prediction shown in Figure~\ref{fig:trustworthypred}.
It is made by SVulD~\cite{Ni:2023:DistinguishingLookAlikeInnocentandVulnerableCodebySubtleSemanticRepresentationLearningandExplanation} for 
the same function \textcode{vrrp\_print\_data}.
This prediction correctly annotates the vulnerable line, Line~7 (32\%), as the most important suspicious line.
Specifically, it highlights the call to \textcode{fopen}, a sensitive operation that can trigger the vulnerability. 
The program attempts to access a file based on the filename \textcode{dump\_file}.
However, it does not properly prevent \textcode{dump\_file} from identifying a link 
that resolves to an unintended resource.
Line~3 (13\%) is also relevant to the vulnerability at Line~7 through a control dependency.
This line is ranked among the top three suspicious lines annotated by the detector.
As a result, the vulnerability prediction made by SVulD is considered trustworthy.}
It is noteworthy that this study focuses on correct predictions only, as incorrect ones can annotate meaningless code parts.

Existing studies~\cite{Qinghua:2022:TowardsARoadmapOnSoftwareEngineeringForResponsibleAI} have acknowledged the use of ML explanations for trustworthiness assessment by disclosing the decision-making process.
These explanations are categorised as global~\cite{Caruana:2015:IntelligibleModelsforHealthCare:PredictingPneumoniaRiskandHospital30dayReadmission} and local~\cite{Ying:2019:GNNExplainer:GeneratingExplanationsforGraphNeuralNetworks, Ribeiro:2016:WhyShouldITrustYouExplainingthePredictionsofAnyClassifier, Li:2017:UnderstandingNeuralNetworksthroughRepresentationErasure}.
The former provides insights into the entire model’s workings.
The latter breaks down a model into its components and focuses on individual predictions.
This allows users to grasp the decision-making process in a way that aligns with human cognition.
Hence, prior studies~\cite{Fu:2022:LineVul:ATransformerBasedLineLevelVulnerabilityPrediction, Ying:2019:GNNExplainer:GeneratingExplanationsforGraphNeuralNetworks} have leveraged the latter for fine-grained interpretations of vulnerability predictions at the line level.
Such interpretations take the form of a list of suspicious lines, each paired with an \textit{importance score}, as described in Table~\ref{table:glossary}.


This study differs from local explanations~\cite{Chu:2024:GraphNeuralNetworksforVulnerabilityDetectionACounterfactualExplanation, Hu:2023:InterpretersforGNNBasedVulnerabilityDetectionAreWeThereYet}. 
They aim to generate faithful explanations that accurately describe model reasoning. 
Conversely, we focus on automatically evaluating the reasonableness of these explanations.
Local explanations are complementary to \truevul{} rather than interchangeable. 
\truevul{} can work with any local explanations, as they offer different interpretations for the same prediction.
We further discuss this in Section~\ref{sec:related_work}.

\begin{figure}[t]
    \centering
    \subfloat[LineVul~\cite{Fu:2022:LineVul:ATransformerBasedLineLevelVulnerabilityPrediction}]{\includegraphics[scale=0.7, clip, trim=2mm 3mm 2mm 0mm]{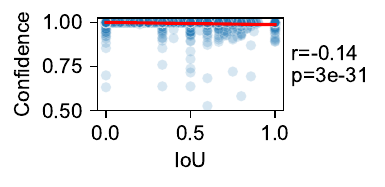}}
    \hspace{4mm}
    \captionsetup[subfigure]{oneside,margin={-8mm,0cm}}
    \subfloat[SVulD~\cite{Ni:2023:DistinguishingLookAlikeInnocentandVulnerableCodebySubtleSemanticRepresentationLearningandExplanation}]{\includegraphics[scale=0.7, clip, trim=15mm 3mm 2mm 0mm]{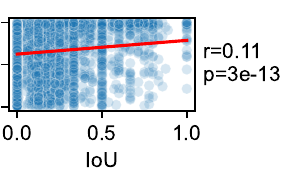}}
    \vspace{-3mm}
    \captionsetup[subfigure]{oneside,margin={0cm,0cm}}
    \subfloat[IVDetect~\cite{Li:2021:VulnerabilityDetectionwithFineGrainedInterpretations}]{\includegraphics[scale=0.7, clip, trim=2mm 3mm 2mm 0mm]{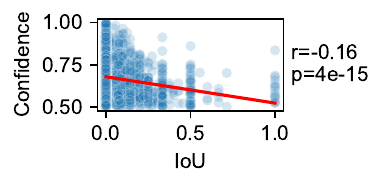}}
    \hspace{4mm}
    \captionsetup[subfigure]{oneside,margin={-10mm,0cm}}
    \subfloat[ReVeal~\cite{Chakraborty:2022:DeepLearningBasedVulnerabilityDetectionAreWeThereYet}]{\includegraphics[scale=0.7, clip, trim=15mm 3mm 2mm 0mm]{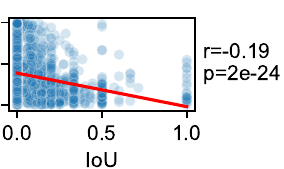}}
    \caption{The distribution of model confidence across Intersection over Union (IoU) between prediction-annotated suspicious lines and ground-truth vulnerable lines of code. Red lines illustrate the trend of Confidence-IoU correlation, with Spearman correlation coefficients and p-values shown.}
    \label{fig:correlation_trust_conf}
\end{figure}

\subsection{Motivation}

\subsubsection{Conventional metrics as insufficient indicators of trustworthiness}

ML models have struggled to pinpoint vulnerabilities,
often due to spurious correlations learnt during training~\cite{Ganz:2021:ExplainingGraphNeuralNetworksforVulnerabilityDiscovery, Chakraborty:2022:DeepLearningBasedVulnerabilityDetectionAreWeThereYet, Steenhoek:2023:AnEmpiricalStudyofDeepLearningModelsforVulnerabilityDetection, Cheng:2024:BeyondFidelityExplainingVulnerabilityLocalizationofLearningBasedDetectors}.
This highlights the need to evaluate not only the overall detection ability using conventional metrics like model confidence or accuracy, but also their 
decision-making.

We conduct a preliminary study showing that conventional metrics like model confidence are insufficient indicators of trustworthiness. 
We collect vulnerability predictions by four SOTA detectors, LineVul~\cite{Fu:2022:LineVul:ATransformerBasedLineLevelVulnerabilityPrediction}, SVulD~\cite{Ni:2023:DistinguishingLookAlikeInnocentandVulnerableCodebySubtleSemanticRepresentationLearningandExplanation}, IVDetect~\cite{Li:2021:VulnerabilityDetectionwithFineGrainedInterpretations}, and ReVeal~\cite{Chakraborty:2022:DeepLearningBasedVulnerabilityDetectionAreWeThereYet}, on BigVul~\cite{Fan:2020:BigVul:ACCCodeVulnerabilityDatasetwithCodeChangesandCVESummaries} dataset.
Their statistics are shown in 
Table~\ref{table:models_under_test}.
Trustworthiness is measured based on 
the overlap between 
suspicious lines ($E$) annotated by the prediction and ground-truth vulnerable lines ($G$).
We use IoU to do this as described in Table~\ref{table:glossary}.
A high IoU indicates that the prediction more 
accurately annotate vulnerable lines, i.e., higher trustworthiness.
To measure model confidence, we use the probability 
calculated by the models that a function is vulnerable.

Figure~\ref{fig:correlation_trust_conf} shows confidence distributions across IoU values.
Across models, untrustworthy predictions (low IoU) are more common than trustworthy ones (high IoU).
Predictions with similar confidence levels can vary in trustworthiness, with both high and low IoU values observed.
Hence, relying on these metrics easily overlooks untrustworthy high-confidence predictions and trustworthy low-confidence predictions.
This limits their ability to detect untrustworthy predictions.
Confidence-IoU correlations represented by the red lines are consistently weak across all models.
Spearman correlation coefficients are all near 0.
SVulD's $r=0.11$ while others' range from $-0.19$ to $-0.14$ ($\text{p-value} < 0.05$).
These statistical results suggest little to no relationship between model confidence and trustworthiness.
Changes in one variable do not reliably predict changes in the other.

\ehl{
This phenomenon is also observed in prior work~\cite{Lam:2024:AutomatedTrustworthinessOracleGenerationforMachineLearningTextClassifiers, Rahman:2024:TowardsCausalDeepLearningforVulnerabilityDetection}. 
ML models are trained on a specific dataset, and their confidence scores are typically derived from \textit{softmax} probabilities.
These probabilities reflect how frequently certain patterns appear in the training data.
They do not indicate whether the patterns steering the prediction are truly correct.
Model confidence is calibrated to the training distribution.
Hence, a model may be highly confident even when relying on spurious patterns 
instead of truly causal ones.
In contrast, the IoU used in this study measures the overlap between the patterns the model relies on and the ground-truth vulnerable features. 
Consequently, model confidence and IoU diverge.
Model confidence quantifies internal probabilistic certainty, while IoU evaluates whether that certainty is grounded in causal and semantically appropriate reasoning.
}

\begin{custombox}[Takeaway~\#1]
SOTA models often make untrustworthy predictions, even with high confidence.
Relying on model confidence can lead to disregarding trustworthy low-confidence predictions while accepting untrustworthy high-confidence ones.
Hence, model confidence is inadequate to detect untrustworthy predictions. 
\end{custombox}

\begin{figure}[t]
    \centering
    \subfloat[Traditional\label{fig:vd_process_wo_truevul}]{\includegraphics[scale=0.69, trim={1mm 1mm 1mm 0mm}, clip]{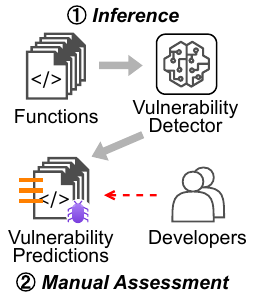}}
    \hfill
    \subfloat[With \truevul\label{fig:vd_process_w_truevul}]{\includegraphics[scale=0.69, trim={1mm 1mm 1mm 0mm}, clip]{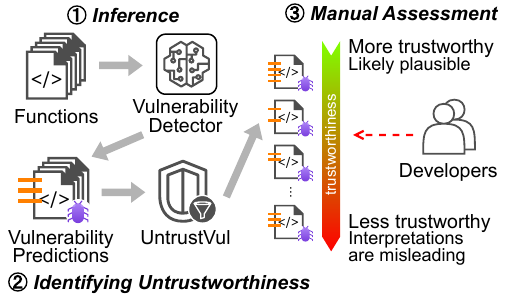}}
    \caption{Two processes of ML-based vulnerability detection.}
\end{figure}
\subsubsection{The role of \truevul{} in ML-based vulnerability detection}
Traditionally, developers need to manually review predicted vulnerabilities by assessing the suspicious lines, as shown in Figure~\ref{fig:vd_process_wo_truevul}.
Unfortunately, Figure~\ref{fig:correlation_trust_conf} shows that many predictions are actually untrustworthy, relying on irrelevant lines.
They can mislead and hinder developers from locating the actual root causes of vulnerabilities, increasing the burden of manual review.
Worse, if developers rely on such suspicious lines, they may create flawed patches that fail to fix true vulnerabilities and even introduce new ones.
At the same time, it is also important not to incidentally exclude any trustworthy predictions, leaving the vulnerabilities unattended.

\textit{\truevul{} is essential for exposing untrustworthy vulnerability predictions during inference}, as using model confidence is unreliable.
Figure~\ref{fig:vd_process_w_truevul} illustrates the vulnerability detection process integrated with \truevul.
To ensure its helpfulness, it is designed to fit seamlessly into developers' workflow.
\truevul{} acts as a filter between vulnerability predictions and developers, exposing untrustworthy predictions.
By doing so, it considers the same input as developers in the traditional process; that is, individual vulnerability predictions that annotate suspicious lines.
Developers can now recognise untrustworthy predictions, decreasing the willingness to rely on them.
They can adjust the efforts invested to review these predictions or even discard them.
Conversely, they can also identify trustworthy predictions,
reducing the efforts of manual assessment.

\section{\truevul}
\label{sec:truevul}


We take as input a vulnerability prediction that annotates suspicious lines and the corresponding source code.
The key idea is to assess whether these suspicious lines are vulnerability-irrelevant.
The prediction is deemed untrustworthy if 
they are mainly
irrelevant.
A line is considered irrelevant if it \ding{172}~is impossible to be vulnerable, i.e., a benign candidate, and \ding{173}~does not reach other suspicious lines that are non-benign candidates, i.e., likely to be vulnerable.
We consider non-benign candidates as a practical estimation of actual vulnerabilities.
These non-benign candidates are potentially vulnerable, which may or may not be the actual vulnerabilities, depending on their context. 
Then, benign candidates that do not even reach non-benign ones are obviously 
irrelevant.

Inspired by prior studies~\cite{Nong:2024:VGX:LargeScaleSampleGenerationforBoostingLearning-BasedSoftwareVulnerabilityAnalyses, Chakraborty:2022:DeepLearningBasedVulnerabilityDetectionAreWeThereYet, Li:2021:VulnerabilityDetectionwithFineGrainedInterpretations}, we assess \ding{172}~by determining whether individual lines violate patterns of vulnerable lines.
Traditional static analysis tools like Checkmarx~\cite{Checkmarx} require manually defined rules, limiting their scalability and adaptability to diverse vulnerability patterns.
Hence, we leverage an ML approach to learn the patterns of vulnerable lines and predict whether a line exhibits different patterns.
The patterns of vulnerable lines have been extensively documented in historical data.
However, this task remains challenging due to the inherently complex and dynamic nature of code syntax and semantics~\cite{Zhou:2025:LargeLanguageModelforVulnerabilityDetectionandRepair:LiteratureReviewandtheRoadAhead, Li:2022:SySeVR:AFrameworkforUsingDeepLearningtoDetectSoftwareVulnerabilities}.
For that reason, we only focus on clearly distinguishable non-vulnerable lines and aim to optimise this goal.

We encode each line into a deep representation using transformer-based code models and compare it with deep representations of historical vulnerable lines.
We frame this comparison as a binary classification task and apply supervised learning to determine whether a line resembles historical non-vulnerable or vulnerable lines.
Lines classified as resembling historical non-vulnerable lines are called \textit{benign candidates}.

Vulnerabilities are context-dependent; thus, some benign candidates can contribute to vulnerabilities via control or data dependencies. 
Heuristic~\ding{173} is introduced to further refine the vulnerability irrelevance of the benign candidates. 
Benign candidates that do not reach any non-benign candidates are obviously irrelevant. 
Predictions relying on them are definitely untrustworthy.
Following this intuition, to assess~\ding{173}, we evaluate whether there is a sequence of dependencies from the line to a non-benign candidate.
Particularly, there exist control or data dependencies that manipulate a variable involved in the non-benign candidate.
As dependencies accumulate in longer sequences, the influence of the line 
becomes less direct, since more intermediate steps can attenuate or override its influence.
Hence, the shorter the sequence, the stronger the influence, meaning the two lines are more related.
We define the length of this sequence as the \textit{reachability distance}, which quantifies the relatedness between two lines.
Finally, a score, denoted by $\mathcal{T}$, is calculated by aggregating the importance scores and reachability distances of the suspicious lines.
\textit{A lower $\mathcal{T}$ indicates the prediction is less likely trustworthy}.

\begin{figure}[t]
    \centering
    \includegraphics[width=0.95\linewidth, clip, trim=0mm 0mm 0mm 1mm]{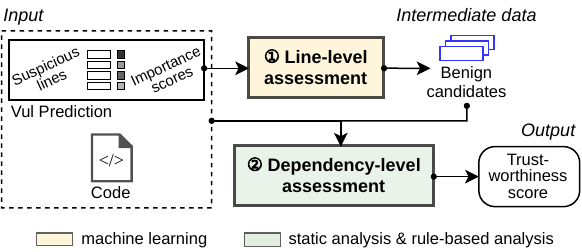}
    \caption{\color{revisedcolor}\truevul's workflow consists of two main stages. 
    It takes as input the source code and the vulnerability prediction that annotates suspicious lines.
    Stage~\ding{172}, line-level assessment, identifies benign candidates among the suspicious lines.
    Stage~\ding{173}, dependency-level assessment, analyses the reachability between the suspicious lines and the non-benign candidates to compute a trustworthiness score.
    }
    \label{fig:truevul}
\end{figure}
Figure~\ref{fig:truevul} shows \truevul's workflow with two stages: line-level assessment (for \ding{172}) and dependency-level assessment (for \ding{173}).
\truevul{} takes as input the \textit{source code} and its vulnerability prediction.
The prediction annotates \textit{suspicious lines} with \textit{importance scores} that quantify their contributions.
This can be represented as $E = \{ \langle l_i, s_i \rangle \}$, where $l_i$ is a line, and $s_i$ indicates its importance score.  
Line-level assessment then categorises these \textit{suspicious lines} into \textit{benign candidates and non-benign ones}.
Next, dependency-level assessment parses the function's \textit{source code} into a program dependency graph (PDG).
It is simplified and combined with 
\textit{importance scores} and \textit{line-level categories} to produce a graph called \textit{weighted PDG}.
In the weighted PDG, each node represents a line with an associated importance score, and each edge denotes a control or data dependency between lines.
Finally, a \textit{score} is calculated by reachability analysis on the \textit{weighted PDG} to estimate the prediction's trustworthiness.
Section~\ref{sec:method_example} 
first presents an example of how \truevul{} processes a prediction, illustrating its two-stage workflow.
Sections~\ref{sec:syntactic_checking} and~\ref{sec:dependency_checking} then describe each stage in detail, respectively.

{\color{revisedcolor}\subsection{Illustrative Example} \label{sec:method_example}

We illustrate how \truevul{} works through its two stages using the prediction made by LineVul shown in Figure~\ref{fig:intro}.
Its interpretation is as follows: $E = \{ \langle L1, 0.13 \rangle,$ $ \langle L2, 0.01 \rangle, \langle L3, 0.06 \rangle, \langle L4, 0.18 \rangle, \langle L5, 0.27 \rangle, \langle L6, 0.02 \rangle, \langle L7,$ $0.08 \rangle, \langle L8, 0.19 \rangle, \langle L9, 0.04 \rangle, \langle L10, 0.01 \rangle, \langle L11, 0.01 \rangle \}$.

In the \textit{first stage}, each suspicious line in $E$ is classified as either a benign or a non-benign candidate.
Specifically, it is transformed into a deep representation using pretrained} {\color{revisedcolor}transformer-based code models.
The resulting representation is then compared with those of historical lines to check if the suspicious line resembles vulnerable or non-vulnerable lines.
We will describe this step in detail in Section~\ref{sec:syntactic_checking}.
In this example, we assume that \truevul{} identifies all suspicious lines as benign candidates, except for Line~7 identified as a non-benign candidate likely to be vulnerable.

In the \textit{second stage}, \truevul{} uses static analysis tools to parse \textcode{vrrp\_print\_data} into a PDG.
The original PDG is simplified by merging nodes from the same line of code into one and preserving their inter-line dependencies. 
Figure~\ref{fig:cpg_example} shows the simplified PDG $G$, where nodes represent lines and edges denote data or control dependencies between them. 
\begin{figure}[t]
    \centering
    \includegraphics[scale=0.86, clip, trim=0 0 1mm 0]{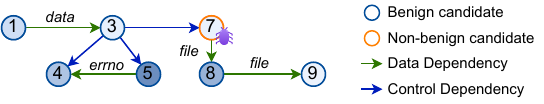}
    \caption{\color{revisedcolor}The PDG of \textcode{vrrp\_print\_data}. 
    Blue and green edges represent control and data dependencies, respectively. 
    Each node corresponds to a line of code, labelled with the line number shown in Figure~\ref{fig:intro}. 
    The outline color indicates whether the line is classified as a benign candidate. 
    The fill color represents the line’s contribution to the prediction, as shown on the right side of Figure~\ref{fig:intro}, where darker shades denote more significant contributions.}
    \label{fig:cpg_example}
\end{figure}
\begin{figure}[t]
    \centering
    \includegraphics[scale=0.86, clip, trim=0 1mm 0mm 0]{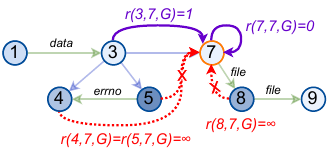}
    \caption{\color{revisedcolor}The analysis and calculation of reachability distances on the PDG of \textcode{vrrp\_print\_data}. 
    For example, \ding{175}, \ding{176}, and \ding{179} cannot reach \ding{178}.
    In contrast, \ding{174} can reach \ding{178} through a control dependency.
    }
    \label{fig:reachability_distance_example}
\end{figure}
\truevul{} next attaches importance scores from $E$ to the corresponding nodes in $G$: 0.13$\rightarrow$\ding{172}, 0.06$\rightarrow$\ding{174}, 0.18$\rightarrow$\ding{175}, 0.27$\rightarrow$\ding{176}, 0.08$\rightarrow$\ding{178}, 0.19$\rightarrow$\ding{179}, 0.04$\rightarrow$\ding{180}.
Importance scores of other lines are ignored, as they are not represented in $G$.

\truevul{} then performs reachability analysis on $G$, as illustrated in Figure~\ref{fig:reachability_distance_example}.
We now compute the reachability distance for each node.
This is the length of the dependency sequence from a node to the nearest non-benign candidate, which is \ding{178}.
\ding{178}'s distance to itself is 0, making it vulnerability-relevant.
Nodes~\ding{175}, \ding{176}, \ding{179}, and \ding{180} cannot reach \ding{178} through any dependency.
Their reachability distances are therefore $\infty$, and they are completely vulnerability-irrelevant.
Node~\ding{174} can reach \ding{178} directly through a single control dependency 3$\rightarrow$7, giving it a reachability distance of 1.
Node~\ding{172} can reach \ding{178} through a sequence: the data dependency 1$\rightarrow$3 and the control dependency 3$\rightarrow$7.
However, the dependency 1$\rightarrow$3 manipulates \textcode{data} that is not used at Line~7. 
Consequently, \ding{172} is considered as vulnerability-irrelevant with a reachability distance of $\infty$.

After reachability analysis, only \ding{174} and \ding{178} have non-infinite reachability distances.
\truevul{} then calculates a score $\mathcal{T}$ to estimates the prediction's trustworthiness.
Intuitively, each suspicious line is associated with an importance score (described in Table~\ref{table:glossary}), the reachability distance to the nearest non-benign candidate, and that candidate's importance score.
The importance scores of the suspicious line and its nearest non-benign candidate positively correlate with the prediction's trustworthiness.
In contrast, the reachability distance correlates negatively, as a smaller value indicates higher vulnerability-relatedness.
To capture this behaviour, we compute $\mathcal{T}$ by aggregating the ratios of the combined importance scores of each suspicious line and its nearest non-benign candidate to their reachability distance.
}
Formally, this score is defined in Equation~\ref{eq:trust_score}, which will be discussed in Section~\ref{sec:trust_score}.
In this example, $\mathcal{T} = \frac{0.06 + 0.08}{2} + \frac{0.08}{1} = 0.15$.
A low value of $\mathcal{T}$ indicates that the prediction is likely untrustworthy, as a large proportion of the suspicious lines are vulnerability-irrelevant.

\subsection{Line-level Assessment for Identifying Benign Candidates} \label{sec:syntactic_checking}

\begin{figure*}[t]
    \centering
    \includegraphics[width=\linewidth, clip, trim=0 3mm 0 0]{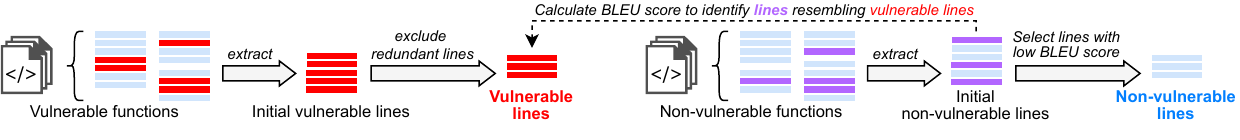}
    \caption{Dataset construction in line-level assessment.}
    \label{fig:data_collect}
\end{figure*}

This step aims to identify benign candidates.
They are lines whose syntax clearly indicates that they are distinguishable non-vulnerable lines and are impossible to be vulnerable.
Our goal is to optimise this process, as perfectly differentiating non-vulnerable lines from vulnerable ones is challenging.
We apply supervised learning to classify a line as a benign candidate or not, training
classifiers using historical vulnerable and non-vulnerable lines.
Worth mentioning, line-level assessment differs from the original vulnerability detection task, which considers each function as a whole.
Models for the original task must comprehend all lines within their contexts.
In contrast, line-level assessment does not focus on function-level vulnerability detection, but determines if each suspicious line annotated by the detector is clearly non-vulnerable
in a context-free manner, while its broader context is assessed in dependency-level assessment.

Many SOTA models~\cite{Cao:2024:Coca:ImprovingandExplainingGraphNeuralNetworkBasedVulnerabilityDetectionSystems,  Fu:2022:LineVul:ATransformerBasedLineLevelVulnerabilityPrediction, Hin:2022:Linevd:StatementLevelVulnerabilityDetectionUsingGraphNeuralNetworks, Li:2021:VulnerabilityDetectionwithFineGrainedInterpretations} have been trained on BigVul~\cite{Fan:2020:BigVul:ACCCodeVulnerabilityDatasetwithCodeChangesandCVESummaries}, a well-known vulnerability dataset.
Hence, we leverage its training set as historical data, followed by multiple cleaning steps to ensure \truevul's robustness.
Specifically, we extract historical vulnerable and non-vulnerable lines from BigVul's training set, as described in Section~\ref{sec:line_level_data_process}.
To improve the effectiveness of line-level assessment, we employ ensemble learning~\cite{Leon:2017:EvaluatingTheEffectOfVotingMethodsOnEnsemble-basedClassification}, aggregating multiple classifiers, as described in Section~\ref{sec:ensemble_learning}.
In Section~\ref{sec:Datasets}, we further evaluate \truevul{} on more recent, high-quality datasets to assess its generalisability to unseen vulnerabilities.

\ehl{
Although \truevul{} relies on historical data, it can generalise effectively due to four principled design choices.
First, most software source code is not entirely new but largely reused, adapted, or combined from historical code. 
Empirical studies show that most vulnerabilities are recurring or highly resemble past ones~\cite{Cao:2025:RecurringVulnerabilityDetection:HowFarAreWe, Pham:2010:DetectingRecurringandSimilarSoftwareVulnerabilities}, while truly novel vulnerabilities are rare, accounting for only 0.1\%~\cite{ZeroDay}.
Our historical data contains 26K vulnerable lines from 2,706 functions, covering 91 CWEs.
This scale and diversity allow \truevul{} to capture the patterns of vulnerable lines.
Second, line-level assessment is inherently simpler and more localised than function-level detection.
Hence, the learnt patterns can be more general and remain stable across vulnerability types.
Third, line-level assessment is deliberately designed to classify lines that contain sensitive operations (e.g., memory allocations, memory accesses, or arithmetic expressions) as non-benign candidates. 
This safe fallback ensures that when encountering unseen data, \truevul{} avoids the critical error of marking untrustworthy predictions as trustworthy.
It can be cautious with unfamiliar patterns, but it does not fail catastrophically. 
Finally, the second stage, dependency-level assessment, analyses reachability on PDGs, which is vulnerability-agnostic.
As a result, \truevul{} can generalise beyond historical data, rather than learning brittle mappings from coarse-grained functions to vulnerabilities.
}

\subsubsection{\textbf{Historical data construction}} \label{sec:line_level_data_process}

Figure~\ref{fig:data_collect} describes the process of dataset construction for training classifiers in line-level assessment.
We leverage the same training set as prior studies~\cite{Li:2021:VulnerabilityDetectionwithFineGrainedInterpretations, Fu:2022:LineVul:ATransformerBasedLineLevelVulnerabilityPrediction} of BigVul~\cite{Fan:2020:BigVul:ACCCodeVulnerabilityDatasetwithCodeChangesandCVESummaries}.
It provides non-vulnerable and vulnerable functions with detailed line-level fixes within git commits that reflect the changes addressing vulnerabilities. 
BigVul was constructed by utilising and linking the CVE (Common Vulnerabilities and Exposures) database, the vulnerability reports, and the code commits.
This helps to improve the accuracy of identifying the vulnerability-related commits with code changes~\cite{Fan:2020:BigVul:ACCCodeVulnerabilityDatasetwithCodeChangesandCVESummaries}.

\textbf{Data cleaning}. Studies~\cite{Croft:2023:DataQualityforSoftwareVulnerabilityDatasets, Ding:2024:VulnerabilityDetectionwithCodeLanguageModelsHowFarAreWe} have reported the existence of tangled commits that involve changes unrelated to vulnerabilities.
This potentially impacts the accuracy of labels in historical data.
To mitigate this problem, we leverage existing studies~\cite{Wang:2024:ReposVul:ARepositoryLevelHighQualityVulnerabilityDataset} that use an LLM, e.g., Qwen\footnote{https://www.alibabacloud.com/en/solutions/generative-ai/qwen} and several open-source static analysis tools, including Cppcheck\footnote{http://cppcheck.sourceforge.net/}, 
Flawfinder\footnote{https://dwheeler.com/flawfinder}, 
RATS\footnote{https://github.com/andrewd/rough-auditing-tool-for-security}, 
and Semgrep\footnote{https://semgrep.dev/}, 
to filter out unrelated changes.
This step does not constitute the novelty of our work; rather, we follow the procedure proposed by~\cite{Wang:2024:ReposVul:ARepositoryLevelHighQualityVulnerabilityDataset}.
\ehl{We provide the LLM 
with the function, the code changes, and CVE supplementary information: the vulnerability description and the commit message.
We then prompt it to evaluate the relevance between the code changes and the vulnerability based on explicit mentions in the supplementary information~\cite{Ding:2024:VulnerabilityDetectionwithCodeLanguageModelsHowFarAreWe, Wang:2024:ReposVul:ARepositoryLevelHighQualityVulnerabilityDataset}. 
Since human experts have analyzed the CVE records, the supplementary information is a reliable reference.
We apply this procedure to 8,736 vulnerable functions in the training set of BigVul.
This results in 4,695 functions.
We later employ the static analysis tools for vulnerability checking. 
The code changes are related if vulnerabilities detected by these tools 
correspond to code changes in the patch~\cite{Wang:2024:ReposVul:ARepositoryLevelHighQualityVulnerabilityDataset}. 
This step leads to 2,706 functions and their deemed related code changes.
These related code changes are then used to extract historical data.
}

\textbf{Extracting historical data}.
Following prior research~\cite{Chu:2024:GraphNeuralNetworksforVulnerabilityDetectionACounterfactualExplanation, Fan:2020:BigVul:ACCCodeVulnerabilityDatasetwithCodeChangesandCVESummaries, 
Hin:2022:Linevd:StatementLevelVulnerabilityDetectionUsingGraphNeuralNetworks, 
Hu:2023:InterpretersforGNNBasedVulnerabilityDetectionAreWeThereYet, 
Li:2021:VulnerabilityDetectionwithFineGrainedInterpretations}, we adopt lines deleted or modified
and lines that are control or
data dependent on the added lines as historical vulnerable lines.
We carefully exclude redundant lines such as comments, blank lines, delimiters, and keyword-only.
The final list of historical vulnerable lines is denoted by $L^+$.
Next, we identify a list of non-vulnerable lines $L^-$.
Just based on syntax, certain non-vulnerable lines may resemble vulnerable lines.
Hence, we aim to construct $L^-$ that is syntactically distinguishable from $L^+$, but diverse and sufficient to train line-level classifiers.
Specifically, we randomly select lines in non-vulnerable functions as an initial list of non-vulnerable lines, denoted by $L_0^-$.
To filter lines that resemble those in $L^+$, we use a lightweight approach that estimates their resemblance  
using BLEU score~\cite{Papineni:2002:Bleu:AMethodforAutomaticEvaluationofMachineTranslation}.
For each line in $L_0^-$, we calculate its BLEU score~\cite{Papineni:2002:Bleu:AMethodforAutomaticEvaluationofMachineTranslation} to measure the lexical precision of n-grams between the line and $L^+$.
A higher BLEU score indicates greater resemblance to vulnerable lines.
Hence, we only choose lines 
with a BLEU score below $\delta_{\textit{BLEU}}$ as historical non-vulnerable lines.
Increasing $\delta_{\textit{BLEU}}$ retains more lines in the final list $L^-$, but potentially introduces more noise.
Section~\ref{sec:sensitivity_analysis} evaluates the impact of $\delta_{\textit{BLEU}}$ on \truevul.
It also assesses this approach's ability to identify vulnerable lines itself, using 
the proportions of misclassified vulnerable and benign lines, respectively.
After collecting data, a dataset of historical vulnerable and non-vulnerable lines is constructed.

\textbf{Post-processing}. 
We 
handle duplicate and conflict samples following the process of Cheng et al.~\cite{Cheng:2024:BeyondFidelityExplainingVulnerabilityLocalizationofLearningBasedDetectors}.
To minimise the risk of learning spurious correlations, all lines are normalised by removing comments and applying style formatting using clang-format~\cite{Imgrund:2023:BrokenPromisesMeasuringConfoundingEffectsinLearningbasedVulnerabilityDiscovery}.
The processed dataset is then used to train classifiers to identify benign candidates.

\subsubsection{\textbf{Ensemble learning}} \label{sec:ensemble_learning}
We leverage SOTA transformer-based architectures, including CodeBERT~\cite{Feng:2020:CodeBERT:APreTrainedModelforProgrammingandNaturalLanguages}, GraphCodeBERT~\cite{Guo:2020:Graphcodebert:PreTrainingCodeRepresentationsWithDataFlow}, and UniXcoder~\cite{Guo:2022:UniXcoder:UnifiedCrossModalPretrainingforCodeRepresentation}, 
to encode lines into vector representations. 
We then use them to train binary classifiers that identify benign candidates based on their syntax.
These SOTA models, however, can be biased due to their training data, potentially affecting 
line-level assessment. 
To mitigate this, we apply ensemble learning~\cite{Leon:2017:EvaluatingTheEffectOfVotingMethodsOnEnsemble-basedClassification} by aggregating $K$ classifiers, $M\textsubscript{\textit{ens}} = \{M_1, M_2, ..., M_K\}$. 
Each classifier $M_i$ outputs 1 if it labels a line as a benign candidate, and 0 otherwise.
The ensemble model $M\textsubscript{\textit{ens}}$ detects benign candidates by combining the decisions from $M_1, M_2, ..., M_K$
using majority voting~\cite{Leon:2017:EvaluatingTheEffectOfVotingMethodsOnEnsemble-basedClassification},
as follows.
\begin{gather} \label{eq:ensemble}
M\textsubscript{\textit{ens}}(l) = 
\begin{cases}
1, & \text{if } \frac{\sum_{M_i \in M\textsubscript{\textit{ens}}} M_i(l)}{K} \ge 0.5, \\
0, & \text{otherwise}.
\end{cases}
\end{gather}


\subsection{Dependency-level Assessment for Measuring Reachability} \label{sec:dependency_checking}

\ehl{The semantic structure of code, such as control and data dependencies, plays an important role in vulnerability analysis~\cite{Chakraborty:2022:DeepLearningBasedVulnerabilityDetectionAreWeThereYet, Zhou:2019:Devign:EffectiveVulnerabilityIdentificationbyLearningComprehensiveProgramSemanticsviaGraphNeuralNetworks}. 
They have been proven to be one of the most effective tools~\cite{Liu:2024:PretrainingbyPredictingProgramDependenciesforVulnerabilityAnalysisTasks, Yamaguchi:2014:ModelingandDiscoveringVulnerabilitieswithCodePropertyGraphs} to detect vulnerabilities.
For example, to detect a use-after-free\footnote{Use after free (CWE-416): https://cwe.mitre.org/data/definitions/416.html} vulnerability, we need to identify whether the argument of a \textcode{free} call is used subsequently.
This requires analysing data dependencies to capture the use of the argument, and control dependencies to determine whether such use may occur after the \textcode{free} call is executed.}

\ehl{\color{revisedcolor}
One of the main principles of vulnerability detection based on program dependencies was established in early studies, notably by Yamaguchi et al~\cite{Yamaguchi:2014:ModelingandDiscoveringVulnerabilitieswithCodePropertyGraphs} in 2014.
They proposed several reachability analyses between a source and a sensitive line using control and data dependencies.
Control dependencies model the execution order of lines, whereas data dependencies identify all lines that produce variables used by a sensitive line.
We adapt this principle to the context of trustworthiness, analysing the relationships between prediction-annotated suspicious lines.
In this paper, non-benign candidates are treated as sensitive lines likely to be vulnerable.
We then assess whether the suspicious lines are annotated reasonably by applying Yamaguchi et al's principle. 
We define rules inspired by their analyses to assess whether a dependency between two lines is vulnerability-relevant.
Then, we analyse whether a suspicious line can reach a non-benign candidate through such relevant dependencies.
}

In particular, a prediction based mainly on benign candidates should not immediately be considered untrustworthy.
Some may serve as a pivotal point in the control or data flow reaching other lines, which can be the actual vulnerabilities.
We assume that more direct dependencies suggest a higher chance that the prediction is trustworthy.
Hence, \truevul{} verifies whether benign candidates
are related to non-benign candidates,
that is, benign candidates flow to any non-benign candidates via 
dependencies. 
Their dependencies are assessed
via two criteria: \textit{relevant dependencies} and \textit{reachability}.
Relevant dependencies lead to lines that are non-benign candidates.
Reachability measures how closely a benign candidate is connected to a non-benign candidate.
The benign candidate closer to a non-benign candidate suggests that it has a stronger influence on the potential vulnerabilities, therefore, a higher likelihood that the prediction is trustworthy.

\subsubsection{\textbf{Source Code Parsing}}



For static code analysis,
\truevul{} parses the source code into a
PDG~\cite{Ferrante:1987:TheProgramDependenceGraphAndItsUseInOptimization}, which is described in Table~\ref{table:glossary}.
We leverage Joern~\cite{Joern}, a widely adopted static analysis platform~\cite{Zhou:2019:Devign:EffectiveVulnerabilityIdentificationbyLearningComprehensiveProgramSemanticsviaGraphNeuralNetworks, Li:2021:VulnerabilityDetectionwithFineGrainedInterpretations, Chakraborty:2022:DeepLearningBasedVulnerabilityDetectionAreWeThereYet}, to parse the source code.
The generated PDG is simplified by merging nodes from the same line into a single node, preserving inter-line dependencies.
The simplified PDG contains nodes indicating lines and edges representing their dependencies.
\truevul{} then combines the simplified PDG and $E$ into a weighted PDG, denoted by $G$.
It assigns each node the importance score of the corresponding line.

For example, the prediction 
in Figure~\ref{fig:intro} 
illustrates $E$ as $\{\langle L_1, 0.13 \rangle, \langle L_2, 0.01 \rangle, \langle L_3, 0.06 \rangle, \langle L_4, 0.18 \rangle, \langle L_5, 0.27 \rangle, \langle L_6,$ $0.02 \rangle, \langle L_7, 0.08 \rangle, \langle L_8, 0.19 \rangle, \langle L_9, 0.04 \rangle, \langle L_{10}, 0.01 \rangle, \langle L_{11},$ $0.01 \rangle\}$.
The simplified PDG of \textcode{vrrp\_print\_data} is shown in Figure~\ref{fig:cpg_example}, where each node represents a line of code.
Importance scores are attached to these nodes as follows: 0.13$\rightarrow$\ding{172}, 0.06$\rightarrow$\ding{174}, 0.18$\rightarrow$\ding{175}, 0.27$\rightarrow$\ding{176}, 0.08$\rightarrow$\ding{178}, 0.19$\rightarrow$\ding{179}, 0.04$\rightarrow$\ding{180}.
Importance scores of other lines are ignored, as those lines are not represented in the PDG.

\subsubsection{\textbf{Relevant dependencies}}

Dependencies between suspicious lines 
annotated by a prediction
should be relevant and lead to vulnerabilities.
Intuitively, trustworthy models should rely on control and data dependencies that can reach vulnerabilities through relevant code flow and variable changes.
We approximate this by assessing dependencies that reach 
non-benign candidates.
We define \textit{relevant dependencies}, which indicate dependencies with a strong connection to non-benign candidates.
For example, Figure~\ref{fig:cpg_example} shows the PDG of the function \textcode{vrrp\_print\_data} presented in Figure~\ref{fig:intro}.
The dependencies 1$\rightarrow$3 and 3$\rightarrow$7 all lead to the vulnerability at Line~7, while the other dependencies, such as 3$\rightarrow$4, 5$\rightarrow$4, 7$\rightarrow$8, and 8$\rightarrow$9, do not.
If the model overrelies on the latter dependencies, it is likely to make an untrustworthy prediction.

\begin{table}[t]
\caption{Rule-based analysis for checking whether the dependency $d = x$~$\rightarrow$~$y$ is a relevant dependency}
\label{table:rule_based_vul_dep}
\resizebox{\linewidth}{!}{
\setlength\tabcolsep{2.5pt} 
\begin{tabular}{|p{0.9cm}|p{7.75cm}|}
\hline
\textbf{} & \makecell[c]{\textbf{Relevant Dependency Rules}}  \\
\hline
\multirow{2}{*}{\makecell[l]{Control\\depend-\\ency}}
& 
\begin{enumerate}[leftmargin=*, noitemsep, partopsep=0pt,topsep=0pt,parsep=0pt, nosep, nolistsep, before=\vspace{-5pt}]
    \item There exists a sequence of dependencies $y$~$\rightarrow$~$y_1$, $y_1$~$\rightarrow$~$y_2$, ..., $y_n$~$\rightarrow$~$z$
    \item The line $z$ is not a benign candidate, where $M\textsubscript{\textit{ens}}(z) \ne 1$ 
\end{enumerate} \\
\hline
\multirow{4}{*}{\makecell[l]{Data\\depend-\\ency}}
& 
\begin{enumerate}[leftmargin=*, noitemsep, partopsep=0pt,topsep=0pt,parsep=0pt, nosep, nolistsep, before=\vspace{-5pt}]
    \item There exists a sequence of dependencies $y$~$\rightarrow$~$y_1$, $y_1$~$\rightarrow$~$y_2$, ..., $y_n$~$\rightarrow$~$z$
    \item The line $z$ is not a benign candidate, where $M\textsubscript{\textit{ens}}(z) \ne 1$ 
    \item The dependency d manipulates a variable involved at the line $z$
\end{enumerate} 
\\
\hline
\end{tabular}%
}
\end{table}

To determine whether a dependency $d = x$~$\rightarrow$~$y$ in the PDG denoted by $G$ is relevant, \truevul{} analyses $d$'s type with its start and end points using the rules listed in Table~\ref{table:rule_based_vul_dep}.
We denote this rule-based analysis 
as a function $p(d, G)$,
which returns 1 if $d$ satisfies the 
rules and 0 otherwise.
For example, in Figure~\ref{fig:cpg_example}, given that Line~7 is not a benign candidate while other lines are,  $p(1$$\rightarrow$$3, G) = p(3$$\rightarrow$$7, G) = 1$, as they manipulate \textcode{data} and the \textcode{false} branch of the if statement at Line~3.
In contrast, $p(3$$\rightarrow$$4, G) = p(3$$\rightarrow$$5, G) = p(7$$\rightarrow$$8, G) = p(8$$\rightarrow$$9, G) = 0$, as Lines~4,~5,~8, and~9 are benign candidates with no subsequent nodes in $G$.


\subsubsection{\textbf{Reachability}} \label{sec:Reachability}
Given 
suspicious lines annotated by the prediction,
some are identified as benign candidates in line-level assessment.
To assess reachability, we examine whether these benign candidates have relevant control or data dependencies to other suspicious lines.
We define the \textit{reachability distance} between two lines, $l_1$ and $l_n$,
as follows.
\begin{dfn} 
\label{def:sem_dist}
The reachability distance between 
$l_1$ and $l_n$ in the PDG $G$, denoted as $r(l_1, l_n, G)$, is the length of the shortest sequence of relevant dependencies in $G$ connecting them, $l_1$$\rightarrow$$l_2$, $l_2$$\rightarrow$$l_3$, ..., $l_3$$\rightarrow$$l_n$, where $\forall i \in [1, n)$: $p(l_i$$\rightarrow$$l_{i+1}, G)$$=$$1$. If no such sequence exists, $r(l_1, l_n, G)$$=$$\infty$.
\end{dfn}
\noindent
To better illustrate this concept, we revisit the PDG of the function \textcode{vrrp\_print\_data} shown in Figure~\ref{fig:cpg_example}.
The analysis and calculation of reachability distances are illustrated in Figure~\ref{fig:reachability_distance_example}.
Considering the benign candidates at Lines~4--5 and the vulnerable line, Line~7, there is no path connecting them, as they belong to two different branches.
This means $r(4, 7, G) = r(5, 7, G) = \infty$.
Similarly, there is no dependency from Line~8 to Line~7, making  $r(8, 7, G)$$=$$\infty$.
In contrast, $r(7, 7, G)$$=$$0$ and $r(3, 7, G)$$=$$1$, as Line~3 has a relevant control dependency leading to Line~7.

We define the measure of vulnerability-irrelevance of a line $l$ in the PDG $G$, denoted by $h(l, G)$, as follows.
\begin{dfn} 
\label{def:related_lines}
$h(l, G)$ measures the shortest sequence of relevant dependencies from $l$ to any non-benign candidate $l_i$ using $r(l, l_i, G)$.
Specifically, \(h(l, G) = \bigl \langle r(l, l_i, G), s_i \bigr \rangle\), where \(\langle l_i, s_i \rangle = argmin_{\langle l_i, s_i \rangle \in E, \, M\textsubscript{\textit{ens}}(l_i) = 0} r(l, l_i, G) \).
\end{dfn}
We use $argmin$ to identify the non-benign candidate $l_i$ 
that has the shortest reachability distance $r(l, l_i, G)$ from $l$ to $l_i$.
The function $h(l, G)$ then returns the reachability distance $h(l, G)^r = r(l, l_i, G)$ and $l_i$'s important score $h(l, G)^s = s_i$.
A higher value of $h(l, G)^r$ indicates a weaker relatedness.
If $l$ is a non-benign candidate, $h(l, G)^r = h(l, G)^s = 0$.
On the other hand, $l$ is completely vulnerability-irrelevant if $h(l, G)^r = \infty$.

\subsubsection{\textbf{Trustworthiness score computation}} \label{sec:trust_score}
\ehl{Intuitively, 
the more vulnerability-irrelevant the annotated suspicious lines are, the less trustworthy the prediction becomes.
Each suspicious line $l_i$ is assigned an importance score $s_i$, indicating how the detector relies on it. 
In addition, how vulnerability-irrelevant $l_i$ is can be approximated by the reachability distance $h(l_i, G)^r$, as discussed in Section~\ref{sec:Reachability}.
The line $l_i$ may be a non-benign candidate ($h(l_i, G)^r=0$), a benign candidate that can reach a non-benign candidate ($0<h(l_i, G)^r\neq\infty$), or a benign candidate that cannot reach one ($h(l_i, G)^r=\infty$).}

\ehl{Each suspicious line $l_i$ is associated with its importance score $s_i$, the reachability distance $h(l_i, G)^r$, and the importance score of its nearest reachable non-benign candidate $h(l_i, G)^s$ (if $h(l_i, G)^r$$\neq$$\infty$).
Using only the reachability distance is inadequate. 
While $l_i$ has a strong connection to likely vulnerable lines, it can be neglected by the detector, i.e., having a low importance score.
This still results in an untrustworthy prediction.
It is thus necessary to combine the importance scores and reachability distances to estimate trustworthiness. 
While a lower $h(l_i, G)^r$ indicates that $l_i$ is less likely to be vulnerability-irrelevant, higher $s_i$ and $h(l_i, G)^s$ reflect how much attention the prediction pays to $l_i$ and the related line.}
We define a score, denoted by $\mathcal{T}$, to estimate the trustworthiness
by aggregating the reachability from prediction-annotated suspicious lines with their importance scores, as follows.
\begin{equation} 
\label{eq:trust_score}
\mathcal{T}(E, G) = \sum_{\langle l_i, s_i \rangle \in E, \, h(l_i, G)^r \neq \infty}  \frac{s_i + h(l_i, G)^s}{1+h(l_i, G)^r} 
\end{equation}
For each suspicious line $l_i$ with $h(l, G)^r \neq \infty$, \truevul{} identifies its nearest line in G that is not a benign candidate.
{\color{revisedcolor}
While $s_i$ and $h(l_i, G)^s$ positively correlate with trustworthiness, $h(l_i, G)^s$ correlates negatively.
To balance the importance scores and reachability distance, we combine them by calculating their ratio.
Specifically, \truevul{} sums the importance scores, $s_i + h(l_i, G)^s$, and calculates the reachability distance $h(l_i, G)^r$. 
It then computes the ratio of $s_i + h(l_i, G)^s$ over $1 + h(l_i, G)^r$.
The ``$1+$'' in the denominator calculates the number of nodes in the reachability path.
This normalizes the result and avoids division by zero in cases where $l_i$ is a non-benign candidate (i.e., $s_i + h(l_i, G)^s = s_i$ and $h(l_i, G)^r = 0$).
}
Finally, by summarising all these ratios, \truevul{} yields the trustworthiness score $\mathcal{T}$ for the prediction.
\ehl{It rewards predictions that assign high importance to vulnerability-relevant lines.
In contrast, those that place high importance on irrelevant lines or low importance on relevant lines are penalized.}
\textit{A lower $\mathcal{T}$score indicates that the prediction is more likely untrustworthy}.


\section{Experimental Setup}
\label{sec:expr_setup}

We study the following research questions (RQs):
\begin{enumerate}[label=\textbf{RQ\arabic*.}, leftmargin=2.8\parindent]
    \item How \textit{effective} is \truevul{} in detecting untrustworthy vulnerability predictions?
    \item How \textit{efficient} is \truevul{} in detecting untrustworthy vulnerability predictions?
    \item 
    How do different \textit{design choices} affect \truevul?
    \item 
    How do $\delta_{\textit{BLEU}}$ and $\delta_{\textit{IoU}}$ influence \truevul?
    \item Can \truevul{} \textit{improve} models' trustworthiness?
\end{enumerate}


\subsection{Models under test}
We experiment with four SOTA detectors shown in Table~\ref{table:models_under_test}, which achieve superior performance and have been widely chosen in prior research~\cite{Chakraborty:2024:RevisitingthePerformanceofDeepLearningBasedVulnerabilityDetectiononRealisticDatasets, Cheng:2024:BeyondFidelityExplainingVulnerabilityLocalizationofLearningBasedDetectors, Zhou:2025:LargeLanguageModelforVulnerabilityDetectionandRepair:LiteratureReviewandtheRoadAhead}.
While these detectors have been widely studied and served as subjects for various evaluations~\cite{Chakraborty:2024:RevisitingthePerformanceofDeepLearningBasedVulnerabilityDetectiononRealisticDatasets}, their trustworthiness remains underexplored~\cite{Cheng:2024:BeyondFidelityExplainingVulnerabilityLocalizationofLearningBasedDetectors}, which motivates our choice.

\begin{table*}[t]
\caption{Vulnerability detection models under test and the statistics of their predictions, with percentage of untrustworthy ones when using IoU=0.5 shown inside the brackets}
\label{table:models_under_test}
\centering
\centering
\begin{tabular}{|l|c|c|r|r|r|r|}
\hline
\multirow{2}{*}{\makecell[c]{\textbf{Model}\\\textbf{Under Test}}} & \multirow{2}{*}{\textbf{Model Type}} & \multirow{2}{*}{\makecell[c]{\textbf{Explanation}\\\textbf{Method}}} & \multicolumn{4}{c|}{\textbf{Number of Predictions}}       \\
\cline{4-7} 
\multicolumn{1}{|l|}{} & \multicolumn{1}{l|}{} & \multicolumn{1}{l|}{} & \multicolumn{1}{l|}{\makecell[c]{BigVul}} & \multicolumn{1}{l|}{\makecell[c]{MegaVul}} & \multicolumn{1}{l|}{\makecell[c]{SARD}} & \multicolumn{1}{l|}{\makecell[c]{PrimeVul}}    \\
\hline
LineVul~\cite{Fu:2022:LineVul:ATransformerBasedLineLevelVulnerabilityPrediction}  &  Transformer-based  &   Attention  & 6,912 {(81\%)} & 1,243 {(90\%)} & 45 {(87\%)}   & 1,021 {(89\%)}                \\
SVulD~\cite{Ni:2023:DistinguishingLookAlikeInnocentandVulnerableCodebySubtleSemanticRepresentationLearningandExplanation}  &  Transformer-based     &   Attention                            & 4,091 {(93\%)} & 10,729 {(88\%)}  & 60,187 {(91\%)}     &   2,323 {(92\%)}          \\
IVDetect~\cite{Li:2021:VulnerabilityDetectionwithFineGrainedInterpretations}  &  Graph-based  &   GNNExplainer      & 2,365  {(64\%)} &   4,694 {(92\%)}  & 4,537 {(89\%)}     & 1,783 {(93\%)}                               \\
ReVeal~\cite{Chakraborty:2022:DeepLearningBasedVulnerabilityDetectionAreWeThereYet}  &  Graph-based  &   GNNExplainer           & 2,716 {(80\%)}  &  5,926 {(90\%)}  & 6,441 {(90\%)}      & 2,053 {(93\%)}                       \\
\hline
\end{tabular}%
\end{table*}

\begin{itemize}
    \item LineVul~\cite{Fu:2022:LineVul:ATransformerBasedLineLevelVulnerabilityPrediction} leverages CodeBERT~\cite{Feng:2020:CodeBERT:APreTrainedModelforProgrammingandNaturalLanguages} to capture long-term dependencies within a long sequence and adopts the attention mechanism to locate the vulnerable lines.
    
    \item SVulD~\cite{Ni:2023:DistinguishingLookAlikeInnocentandVulnerableCodebySubtleSemanticRepresentationLearningandExplanation} adopts contrastive learning to train a UniXcoder~\cite{Guo:2022:UniXcoder:UnifiedCrossModalPretrainingforCodeRepresentation} model to distinguish semantic representations of functions regardless of their lexical similarity.
    
    \item IVDetect~\cite{Li:2021:VulnerabilityDetectionwithFineGrainedInterpretations} represents code as PDGs, employs graph convolution networks with feature attention, and uses GNNExplainer~\cite{Ying:2019:GNNExplainer:GeneratingExplanationsforGraphNeuralNetworks} for line-level interpretations.
    
    \item ReVeal~\cite{Chakraborty:2022:DeepLearningBasedVulnerabilityDetectionAreWeThereYet} translates code into a graph embedding and trains a representation learner on the extracted features.
\end{itemize}

We use the configurations and settings used in the original studies for vulnerability predictions. 
Suspicious lines are identified using the same explanation methods, which are validated for achieving the best performance in fine-grained detection~\cite{Li:2021:VulnerabilityDetectionwithFineGrainedInterpretations, Fu:2022:LineVul:ATransformerBasedLineLevelVulnerabilityPrediction, Zou:2022:mVulPreter:AMultiGranularityVulnerabilityDetectionSystemWithInterpretations}.
For transformer-based models, we leverage the attention mechanism.
For graph-based models, we use GNNExplainer~\cite{Ying:2019:GNNExplainer:GeneratingExplanationsforGraphNeuralNetworks} and apply Joern to parse source code as in the original papers.
%

\subsection{Datasets}
\label{sec:Datasets}

\ehl{
\textit{Dataset selection:}
This study requires datasets with line-level vulnerability annotations. 
We therefore exclude several unsatisfied well-known datasets.
Devign~\cite{Zhou:2019:Devign:EffectiveVulnerabilityIdentificationbyLearningComprehensiveProgramSemanticsviaGraphNeuralNetworks} and Reveal~\cite{Chakraborty:2022:DeepLearningBasedVulnerabilityDetectionAreWeThereYet} are not selected, as they only label vulnerabilities at the function level.
We also exclude D2A~\cite{Zheng:2021:ADatasetBuiltforAI-BasedVulnerabilityDetectionMethodsUsingDifferentialAnalysis}.
Despite providing line-level vulnerable traces, it suffers from a high false positive rate, with over two-thirds of samples mislabeled~\cite{Croft:2023:DataQualityforSoftwareVulnerabilityDatasets}.
}

Consequently, we use BigVul~\cite{Fan:2020:BigVul:ACCCodeVulnerabilityDatasetwithCodeChangesandCVESummaries}, MegaVul~\cite{Ni:2024:MegaVul:AC/CppVulnerabilityDatasetwithComprehensiveCodeRepresentations}, SARD~\cite{SARD}, and PrimeVul~\cite{Ding:2024:VulnerabilityDetectionwithCodeLanguageModelsHowFarAreWe}, which provide line-level vulnerabilities.
BigVul, MegaVul, and PrimeVul contain real-world vulnerabilities, while most vulnerabilities in SARD are synthetic.
BigVul is used for building historical data and intra-project evaluations, as it is used to train SOTA models in prior studies~\cite{Fu:2022:LineVul:ATransformerBasedLineLevelVulnerabilityPrediction, Ni:2023:DistinguishingLookAlikeInnocentandVulnerableCodebySubtleSemanticRepresentationLearningandExplanation, Li:2021:VulnerabilityDetectionwithFineGrainedInterpretations, Chakraborty:2022:DeepLearningBasedVulnerabilityDetectionAreWeThereYet}.
The other datasets that are more recent and distinct from BigVul with high label accuracy~\cite{Croft:2023:DataQualityforSoftwareVulnerabilityDatasets, Ding:2024:VulnerabilityDetectionwithCodeLanguageModelsHowFarAreWe} are used for cross-project evaluations.

\ehl{\textit{Dataset preprocessing:}
For BigVul, we follow prior studies~\cite{Li:2021:VulnerabilityDetectionwithFineGrainedInterpretations, Fu:2022:LineVul:ATransformerBasedLineLevelVulnerabilityPrediction} in using the same partitioning of training, test and validation set.
The training data is used to train the models under test and to construct historical data in line-level assessment.
The testing data is then used for intra-project evaluation.
In addition, following~\cite{Cheng:2024:BeyondFidelityExplainingVulnerabilityLocalizationofLearningBasedDetectors, Ding:2024:VulnerabilityDetectionwithCodeLanguageModelsHowFarAreWe}, we exhaustively compare the vulnerable functions to remove duplicate and conflicting samples both within each dataset as well as between BigVul and the other datasets to prevent data leakage. 
Specifically, we normalize the formatting characters in the code samples to eliminate their noisy effect and then employ a method based on MD5 values to compare two strings after the normalization.}

\textit{Prediction collection:}
We collect predictions made by the models on all datasets.
Our goal is to verify whether \truevul{} correctly judges the trustworthiness of the generated predictions. 
To do so, we need ground-truth labels that indicate whether a vulnerability prediction is trustworthy or not.
We assign the ground-truth trustworthiness label for a prediction by quantifying the overlap between prediction-annotated suspicious lines ($E$) and ground-truth vulnerable lines ($G$).
Specifically, we first perform multiple filtering steps from~\cite{Hin:2022:Linevd:StatementLevelVulnerabilityDetectionUsingGraphNeuralNetworks} to reduce noise in ground-truth vulnerable lines.
This includes removing comments, ignoring cosmetic changes, e.g., whitespace.
We then follow Hu et al.~\cite{Hu:2023:InterpretersforGNNBasedVulnerabilityDetectionAreWeThereYet}, using Intersection over Union (IoU), defined as \(\frac{|E\;\cap\;G|}{|E\;\cup\;G|}\), and consider a prediction untrustworthy if $IoU \le \delta_{\textit{IoU}}$.
We explore the influence of $\delta_{\textit{IoU}}$
in \textbf{RQ4}, while other RQs use the value of 0.5 for both thresholds, as this setting yields the optimal results.
The statistics of the ground truths are outlined in Table~\ref{table:models_under_test}.

\subsection{Metrics}

As our goal is to identify untrustworthy predictions, we frame this task as a binary classification task, with ``untrustworthy'' as the positive class and ``trustworthy'' as the negative class.
We use standard classification metrics, such as Accuracy, Precision, Sensitivity, and F1-score.
However, due to the imbalanced nature of the datasets (Table~\ref{table:models_under_test}), evaluating \truevul's effectiveness is challenging.
Although \truevul{} is not an ML model trained on such imbalanced datasets, it is still important to select appropriate evaluation metrics to avoid misleading results.
Hence, we also include Specificity to assess the ability to detect trustworthy predictions, and metrics like AUC and G-mean to provide a balanced evaluation across both classes.
Specifically, we focus on the following metrics in the main text.
\begin{itemize}
    \item 
    \textit{Accuracy}: the proportion of predictions correctly labeled as untrustworthy or not, calculated as $\frac{TP+TN}{TP+FP+TN+FN}$.
    \item 
    \textit{AUC}: the area under the receiver operating characteristic curve, assessing the ability to distinguish between trustworthy and untrustworthy predictions. 
    \item 
    \textit{F1-score}: the harmonic mean of \textit{Precision} and \textit{Sensitivity}, calculated as $\frac{2 \times \text{\textit{Pre}} \times \text{\textit{Sen}}}{\text{\textit{Pre}} + \text{\textit{Sen}}}$.
    \item 
    \textit{G-mean}: the geometric mean of \textit{Sensitivity} and \textit{Specificity}, calculated as $\sqrt{\text{\textit{Sen}} \times \text{\textit{Spec}}}$.
\end{itemize}
In the supplementary material, we provide detailed results and analysis with additional metrics as follows.
\begin{itemize}
    \item 
    \textit{Precision}: the proportion of true untrustworthy predictions among the detected ones, defined as $\frac{TP}{TP+FP}$.
    \item 
    \textit{Sensitivity} (or Recall): the proportion of untrustworthy predictions correctly detected, defined as $\frac{TP}{TP+FN}$.
    \item 
    \textit{Specificity}: the proportion of correctly detected trustworthy predictions, defined as $\frac{TN}{TN+FP}$.
\end{itemize}
\ehl{To evaluate efficiency, we measure the processing time per function (in seconds) and the average function size (in LOC). 
Because the studied datasets contain real-world vulnerabilities, these metrics also reflect the practical feasibility and applicability of \truevul.
We will further discuss in Section~\ref{sec:Efficiency}.}
To answer \textbf{RQ5}, we report the improved models' vulnerability detection performance using F1-score, and estimate their trustworthiness via the average $\overline{IoU}$ between model-identified suspicious lines and ground-truth vulnerable lines, and the average $\overline{\mathcal{T}}$ score calculated by \truevul.

{\color{revisedcolor}
\subsection{Baselines}

We use the following baselines for comparison.
\begin{enumerate}[leftmargin=*]
\item \textit{Uncertainty-based baselines:}
Many studies~\cite{Zhang:2020:EffectOfConfidenceAndExplanationOnAccuracyAndTrustCalibrationInAiAssistedDecisionMaking, Lam:2024:AutomatedTrustworthinessOracleGenerationforMachineLearningTextClassifiers} have established model uncertainty as a standard trustworthiness proxy in ML.
Accordingly, we include \textit{Naive}, a straightforward approach that detects untrustworthy vulnerability predictions solely based on model certainty. 
Our study uses model confidence as a certainty metric. 
Predictions are deemed untrustworthy if {confidence} $< \theta_{\text{\textit{conf}}}$. 
The optimal value of $\theta_{\text{\textit{conf}}}$ is determined by maximising G-mean.

\item \textit{Perturbation-based baselines:}
Perturbation-based methods are commonly used to reveal spurious correlations by modifying inputs and observing prediction changes.
In vulnerability detection, Rahman et al~\cite{Rahman:2024:TowardsCausalDeepLearningforVulnerabilityDetection} introduced \textit{PerturbVar}, \textit{PerturbAPI}, \textit{PerturbJoint}.
These methods apply semantics-preserving code transformations designed to expose spurious correlations.
Specifically, PerturbVar renames variables to frequently occurring names in the opposite class, PerturbAPI injects frequent API calls from the opposite class, and PerturbJoint combines PerturbVar and PerturbAPI.
Predictions altered by these perturbations are deemed untrustworthy.

\end{enumerate}

Line-level ML detectors can serve as baselines by flagging predictions with minimal overlap with ground-truth vulnerable lines as untrustworthy.
However, this introduces role ambiguity, as they would function simultaneously as both vulnerability detectors and untrustworthiness detectors. 
It also leads to circular evaluation, since different detectors often highlight incorrect lines.
To avoid this, we treat line-level detectors (e.g., LineVul, IVDetect) as models under test.
This enables a clearer experimental design and demonstrates \truevul’s generalisability across detector types.

Regarding existing XAI methods, 
they primarily aim to generate faithful interpretations that correctly describe the model reasoning~\cite{Ying:2019:GNNExplainer:GeneratingExplanationsforGraphNeuralNetworks}.
In contrast, our focus is on trustworthiness through the lens of plausibility, i.e., how reasonable the interpretations are.
In vulnerable detection, it means the prediction highlights lines of code relevant to the actual vulnerability. 
There may be a limited number of studies~\cite{Lam:2024:AutomatedTrustworthinessOracleGenerationforMachineLearningTextClassifiers} in other domains that automatically assess plausibility or prediction trustworthiness. 
However, they do not apply to vulnerability detectors. 
To our knowledge, no existing method can automatically evaluate plausibility in vulnerability detection.

In summary, the selected uncertainty- and perturbation-based approaches
constitute the only applicable approaches as comparator baselines for our study.
}

\begin{table*}
\centering
\caption{The \textit{intra-project} effectiveness of \truevul{} (ours) and the baselines on BigVul dataset}
\vspace{-3mm}
\label{table:expr_results}
\setlength\tabcolsep{2.5pt} 

\subfloat[LineVul]{%
\begin{tabular}{|l|CCCC|}
\hline
             & Acc  & Auc   & F1   & Gm   \\
\hline
\truevul     & 0.78 & 0.86  & 0.86 & 0.80 \\
Naive        & 0.46 & 0.32  & 0.60 & 0.39 \\
PerturbVar   & 0.12 & 0.27  & 0.03 & 0.10 \\
PerturbAPI   & 0.14 & 0.46  & 0.02 & 0.10 \\
PerturbJoint & 0.14 & 0.42  & 0.02 & 0.10 \\
\hline
\end{tabular}%
}
\hspace{1pt}
\subfloat[SVulD]{%
\begin{tabular}{|CCCC|}
\hline
Acc  & Auc   & F1   & Gm   \\
\hline
0.74 & 0.77  & 0.82 & 0.79 \\
0.58 & 0.60  & 0.69 & 0.58 \\
0.44 & 0.56  & 0.52 & 0.51 \\
0.40 & 0.55  & 0.47 & 0.48 \\
0.45 & 0.56  & 0.54 & 0.52 \\
\hline
\end{tabular}%
}
\hspace{1pt}
\subfloat[IVDetect]{%
\begin{tabular}{|CCCC|}
\hline
Acc  & Auc   & F1   & Gm   \\
\hline
0.84 & 0.81  & 0.91 & 0.78 \\
0.32 & 0.32  & 0.47 & 0.38 \\
0.05 & 0.31  & 0.04 & 0.13 \\
0.05 & 0.35  & 0.02 & 0.10 \\
0.06 & 0.32  & 0.04 & 0.14 \\
\end{tabular}%
}
\hspace{1pt}
\subfloat[ReVeal]{%
\begin{tabular}{|CCCC|}
\hline
Acc  & Auc   & F1   & Gm   \\
\hline
0.84 & 0.83  & 0.91 & 0.79 \\
0.36 & 0.31  & 0.51 & 0.38 \\
0.05 & 0.28  & 0.02 & 0.11 \\
0.05 & 0.32  & 0.01 & 0.07 \\
0.05 & 0.30  & 0.02 & 0.10 \\
\hline
\end{tabular}%
}
\end{table*}

\begin{table*}
\centering
\caption{The \textit{cross-project} effectiveness of \truevul{} (ours) and the baselines on MegaVul, SARD, and PrimeVul}
\label{table:expr_results_cp}
\vspace{-3mm}
\setlength\tabcolsep{2.5pt} 

\subfloat[LineVul]{%
\begin{tabular}{|l|CCCC|}
\hline
             & Acc  & Auc   & F1   & Gm   \\
\hline
\truevul     & 0.84 & 0.88  & 0.91 & 0.86 \\
Naive        & 0.38 & 0.56  & 0.64 & 0.56 \\
PerturbVar   & 0.17 & 0.55  & 0.28 & 0.38 \\
PerturbAPI   & 0.08 & 0.77  & 0.12 & 0.23 \\
PerturbJoint & 0.16 & 0.77  & 0.24 & 0.34 \\
\hline
\end{tabular}%
}
\hspace{1pt}
\subfloat[SVulD]{%
\begin{tabular}{|CCCC|}
\hline
Acc  & Auc   & F1   & Gm   \\
\hline
0.80 & 0.84  & 0.86 & 0.81 \\
0.56 & 0.54  & 0.71 & 0.53 \\
0.24 & 0.55  & 0.33 & 0.39 \\
0.20 & 0.54  & 0.28 & 0.37 \\
0.24 & 0.55  & 0.34 & 0.39 \\
\hline
\end{tabular}%
}
\hspace{1pt}
\subfloat[IVDetect]{%
\begin{tabular}{|CCCC|}
\hline
Acc  & Auc   & F1   & Gm   \\
\hline
0.82 & 0.80  & 0.90 & 0.77 \\
0.30 & 0.28  & 0.44 & 0.35 \\
0.03 & 0.26  & 0.02 & 0.09 \\
0.03 & 0.29  & 0.01 & 0.03 \\
0.04 & 0.28  & 0.03 & 0.11 \\
\end{tabular}%
}
\hspace{1pt}
\subfloat[ReVeal]{%
\begin{tabular}{|CCCC|}
\hline
Acc  & Auc   & F1   & Gm   \\
\hline
0.70 & 0.72  & 0.82 & 0.70 \\
0.38 & 0.32  & 0.53 & 0.37 \\
0.03 & 0.31  & 0.01 & 0.05 \\
0.03 & 0.34  & 0.01 & 0.05 \\
0.03 & 0.34  & 0.01 & 0.07 \\
\hline
\end{tabular}%
}
\end{table*}

{\color{revisedcolor}
\subsection{Implementation Details}

\subsubsection*{Hyperparameter settings for line-level assessment}
We explored three pre-trained code models ($K$$=$$3$), CodeBERT~\cite{Feng:2020:CodeBERT:APreTrainedModelforProgrammingandNaturalLanguages}, GraphCodeBERT~\cite{Guo:2020:Graphcodebert:PreTrainingCodeRepresentationsWithDataFlow}, and UniXCoder~\cite{Guo:2022:UniXcoder:UnifiedCrossModalPretrainingforCodeRepresentation}, for ensemble learning. 
For each model, we used 12 transformer encoder blocks, a hidden size of 768, and 12 attention heads. 
We follow the same fine-tuning strategy provided by Feng et al.~\cite{Feng:2020:CodeBERT:APreTrainedModelforProgrammingandNaturalLanguages}. 
During training, we set the number of training epochs to 10 and used the learning rate of 2e-5.
We applied backpropagation with AdamW optimizer~\cite{Ilya:2019:DecoupledWeightDecayRegularization}, which is widely adopted to fine-tune transformer-based models.

\subsubsection*{Static analysis}
We utilized Joern~\cite{Joern} to parse source code into PDGs and stored the resulting graphs as CSV files.

\subsubsection*{Threshold selection}
Our study involves two thresholds, $\delta_{\textit{IoU}}$ and $\delta_{\textit{BLEU}}$.
The former is used to establish ground-truth trustworthiness labels (Section~\ref{sec:Datasets}).
The latter is used to extract historical non-vulnerable lines (Section~\ref{sec:line_level_data_process}).
Their default values are set to 0.5, and we will determine their optimal values through sensitivity analysis (RQ4) in Section~\ref{sec:sensitivity_analysis}.

\subsubsection*{Hardware specifications}
We conducted experiments on a Rocky Linux server with 24 cores of Intel(R) Xeon(R) Gold 6150 CPU @ 2.70GHz, 256GB RAM, and NVIDIA A100 GPU with 40GB memory.
}

\section{Experimental Results}
\label{sec:expr_results}

This section presents the results of experiments evaluating \truevul{} in detecting untrustworthy predictions.

\subsection{\textbf{RQ1: Effectiveness}}

\subsubsection{\textbf{Quantitative analysis}}
Tables~\ref{table:expr_results}--\ref{table:expr_results_cp} show the effectiveness of \truevul{} and the baselines on BigVul dataset (intra-project setting) and on MegaVul, SARD, and PrimeVul datasets (cross-project setting), respectively.
Several trends can be observed from the experiential results.
\begin{itemize}[leftmargin=*]
\item \textit{\truevul{} consistently performs well across both transformer- and graph-based models, and in both intra-project and cross-project settings}.
Specifically, it achieves Accuracy of 70\%--84\%, AUC of 72\%--86\%, F1-score of 82\%--91\%, and G-mean of 70\%--81\%.
This shows \truevul's robustness despite data drifts from the historical data used for ensemble learning in line-level assessment.

\item \textit{Compared to Naive, \truevul{} outperforms in all metrics}, with improvements of 16\%--52\% in Accuracy, 17\%--54\% in AUC, 13\%--45\% in F1-score, and 22\%--67\% in G-mean.
This further demonstrates that model confidence is inadequate to detect untrustworthy predictions.

\item \textit{Compared to PerturbVar, PerturbAPI, and PerturbJoint, \truevul{} improves} by 29\%--82\% in Accuracy, 21\%--59\% in AUC, 28\%--92\% in F1-score, and 27\%--72\% in G-mean.
This may be due to spurious features beyond identifiers, or inadequate perturbations failing to alter predictions.
\end{itemize}

\begin{figure}[t]
\centering
\subfloat[Acc]{\includegraphics[scale=0.55, clip, trim=3mm 5mm 29mm 20mm]{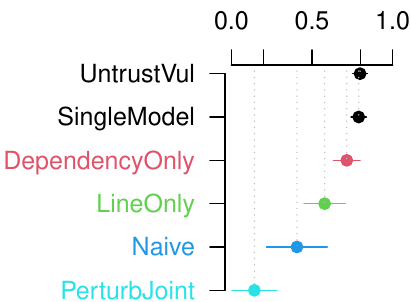}}
\hfill
\subfloat[Auc]{\includegraphics[scale=0.55, clip, trim=3mm 5mm 29mm 20mm]{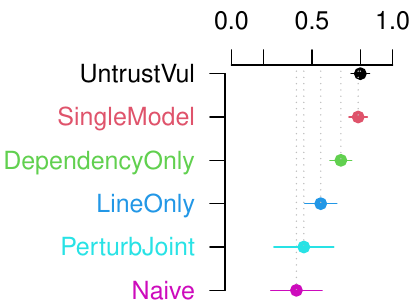}}
\hfill
\subfloat[F1]{\includegraphics[scale=0.55, clip, trim=3mm 5mm 29mm 20mm]{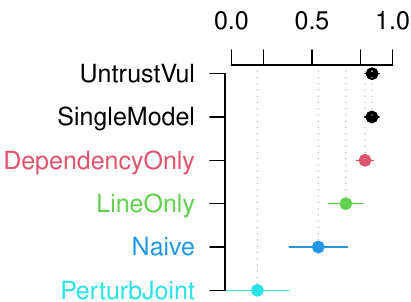}}
\hfill
\subfloat[Gm]{\includegraphics[scale=0.55, clip, trim=3mm 5mm 29mm 20mm]{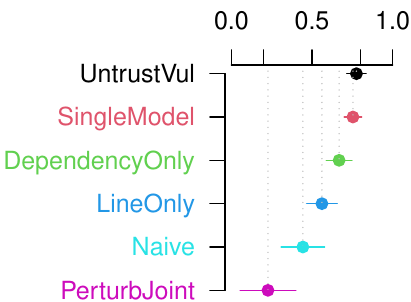}}
\caption{
Comparison of \truevul{}, baselines, and ablations using the Scott-Knott Effect Size Difference (ESD) test~\cite{Tantithamthavorn:2017:AnEmpiricalComparisonofModelValidationTechniquesforDefectPredictionModels}, with colors indicating statistically distinct clusters. ``SingleModel'' represents the average effectiveness of three single-model variants.}
\label{fig:ScottKnottESD_perf}
\end{figure}

We further use the Scott-Knott Effect Size Difference (ESD) test~\cite{Tantithamthavorn:2017:AnEmpiricalComparisonofModelValidationTechniquesforDefectPredictionModels} that leverages hierarchical clustering to partition the set of treatment averages (e.g., means) into statistically distinct groups with non-negligible differences.
Figure~\ref{fig:ScottKnottESD_perf} demonstrates that the superior effectiveness of \truevul{} compared to the Naive and PerturbJoint is statistically significant, as they are denoted by three distinct clusters.

\subsubsection{\textbf{Qualitative analysis}}

{\color{revisedcolor}
Compared with the baselines, \truevul{} detects a total of 40,251 additional untrustworthy predictions. 
Among these predictions, we identify 816 edge cases of untrustworthiness.
Specifically, program dependencies are ambiguous or suspicious lines are not actual vulnerable lines but merely syntactically similar to vulnerable ones.
As a result, \truevul{} would have saved developers from chasing irrelevant lines of code, especially in the 816 edge cases that are difficult to detect and can easily mislead developers.

We also compare \truevul's effectiveness across different vulnerability types.
It typically achieves Accuracy of 75\%–85\%, FPR of 10\%–32\%, and FNR of 15\%–29\% across the top 25 most dangerous software weaknesses in 2025~\cite{CWE:2024:Top25}.

Due to space constraints, we provide detailed results and analyses of several representative TP, TN, FP, and FN cases in the supplementary material.
}

\begin{custombox}[Takeaway~\#2]
\truevul{} consistently outperforms baselines across models and datasets, showing its effectiveness and robustness in exposing untrustworthy predictions.
\end{custombox}

\subsection{\textbf{RQ2: Efficiency}} \label{sec:Efficiency}

\begin{table}[t]
\caption{Comparing the efficiency between approaches}
\label{table:expr_efficiency}
\setlength\tabcolsep{2.5pt} 
\centering
\vspace{-1mm}
\subfloat[Comparing with the baselines]{%
\centering\begin{tabular}[t]{|l|c|c|c|c|c|c|c|c|}
\hline
                & \multicolumn{2}{c|}{BigVul}                         & \multicolumn{2}{c|}{MegaVul}                        & \multicolumn{2}{c|}{SARD}    & \multicolumn{2}{c|}{PrimeVul}                       \\
\cline{2-9}
                & LOC & Time & LOC & Time & LOC & Time & LOC & Time \\
\hline
\truevul{}  & \multirow{4}{*}{147}    & 1.47                     & \multirow{4}{*}{119}    &  1.35                    & \multirow{4}{*}{70}     & 1.02  & \multirow{4}{*}{167}     & 1.52                     \\
PerturbVar      &                         & 0.58                       &                         &  0.56                     &                         &  0.51       & & 0.58                 \\
PerturbAPI      &                         & 0.57                         &                         & 0.56                         &                         & 0.51    & & 0.57                   \\
PerturbJoint    &                         &  0.58                        &                         & 0.57                         &                         & 0.51      & & 0.59                   \\
\hline
\end{tabular}%
\label{tab:time_baseline}
}

\subfloat[Ablation study\label{tab:time_ablation}]{\centering\begin{tabular}[t]{|l|c|c|c|c|}
\hline
                & BigVul                         & MegaVul                        & SARD    & PrimeVul                      \\
\hline
Line only & 1.02 & 0.77 & 0.56 & 1.03 \\
Dependency only        & 0.48               & 0.49                     & 0.44   & 0.48             \\
CodeBERT only          & 0.91                         & 0.60                              & 0.51    &    0.92                \\
GraphCodeBERT only     & 0.93                            & 0.61                            & 0.52    & 0.93                     \\
UniXcoder only            & 0.89                      & 0.60                        & 0.50     &  0.90            \\
\hline
\end{tabular}}
\end{table}

Table~\ref{table:expr_efficiency}\subref*{tab:time_baseline} shows the average lines of code (LOC) per function, and compares the efficiency of \truevul{}, PerturbVar, PerturbAPI, and PerturbJoint by processing time (in seconds) per function. 
\truevul{} takes 1.02--1.47 seconds to assess a prediction, depending on function size.
Compared to the baselines, \truevul{} requires an additional 0.51--0.90 seconds per function due to Joern, the static code analysis tool, and multiple line-level assessment models.

\ehl{
BigVul, MegaVul, and PrimeVul all consist of real-world vulnerabilities. 
Therefore, the results on these datasets also indicate the practical efficiency of \truevul{}.
Given an average size of 350,542 LOC of open-source software reported in prior work~\cite{Woo:2021:CENTRIS:APreciseandScalableApproachforIdentifyingModifiedOpenSourceSoftwareReuse}, \truevul{} only takes approximately 3,190 seconds to scan an entire project on the machine specifications described in Section~\ref{sec:expr_setup}.
This overhead may limit \truevul's applicability in time-sensitive contexts.
However, it is acceptable for security-critical tasks, where deeper analysis is warranted.
}

\begin{custombox}[Takeaway~\#3]
\truevul{} requires only 1.02--1.47 seconds to assess a prediction.
Although it takes more time than the baselines, the increase in time is insignificant given its performance benefits over other approaches.
\end{custombox}

\subsection{\textbf{RQ3: Ablation Study}}

We compare \truevul{} with its ablation variants.
Specifically, the line-only variant considers a prediction untrustworthy if most prediction-contributing lines are classified as benign.
The dependency-only variant randomly selects benign candidates.
We also use three variants using individual models, CodeBERT~\cite{Feng:2020:CodeBERT:APreTrainedModelforProgrammingandNaturalLanguages}, GraphCodeBERT~\cite{Guo:2020:Graphcodebert:PreTrainingCodeRepresentationsWithDataFlow}, and UniXcoder~\cite{Guo:2022:UniXcoder:UnifiedCrossModalPretrainingforCodeRepresentation}, instead of ensemble learning.

Figure~\ref{fig:ScottKnottESD_perf} illustrates results of the Scott-Knott Effect Size Difference test~\cite{Tantithamthavorn:2017:AnEmpiricalComparisonofModelValidationTechniquesforDefectPredictionModels} comparing \truevul{} with its ablations.
We only report the overall and balanced metrics, Accuracy, AUC, F1-score, and G-mean.
Overall, \truevul{} outperforms all ablation variants.
The line-only and dependency-only variants both significantly decrease the effectiveness by 11\%--42\% in Accuracy, 3\%--42\% in AUC, 1\%--37\% in F1-score, and 4\%--42\% in G-mean.
The variants without ensemble learning show relatively lower decreases compared to the other variants.
Ensemble learning overall improves the effectiveness by up to 3\% in G-mean.

Regarding efficiency, Table~\ref{tab:time_ablation}
compares the processing time of \truevul{} and its ablation variants per function.
Static code analysis and multiple line-level assessment models represent most of the processing time, 33\%--69\% of the total.
Ensemble learning 
also affects efficiency, as using a single model reduces the time by 0.50--0.75 seconds.


\begin{custombox}[Takeaway~\#4]
Detecting untrustworthy predictions based solely on syntax or dependencies may miss crucial insights, making it less effective.
\truevul{} overcomes this limitation by integrating insights from both the syntax of individual lines and their dependencies.
Ensemble learning further enhances the effectiveness by up to 3\% in G-mean.
\truevul's efficiency mainly depends on the static code analysis tool, the number of models for line-level assessment, and their efficiency.
\end{custombox}

\subsection{\textbf{RQ4: Sensitivity Analysis}} \label{sec:sensitivity_analysis}

\subsubsection{\textbf{Impact of BLEU thresholds}}

\begin{figure}[t]
    \centering
    \includegraphics[scale=0.5]{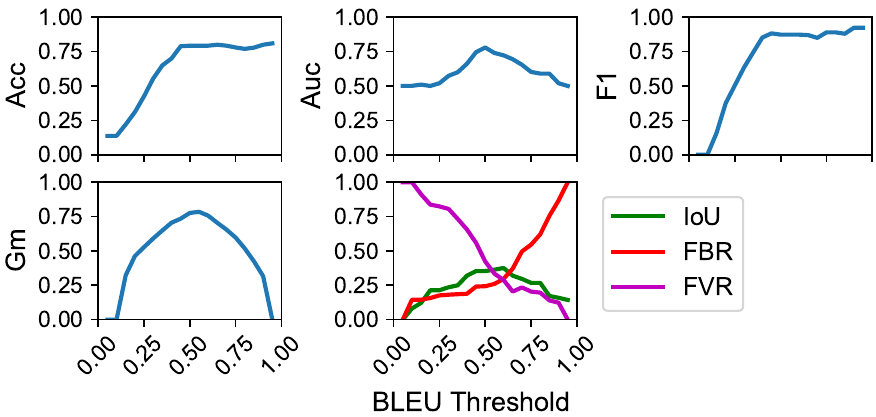}
    \caption{\color{revisedcolor}The impact of the BLEU threshold, $\delta_{\textit{BLEU}}$.
    The first four subplots illustrate the changes in overall effectiveness using Accuracy, AUC, F1-score, and G-mean, respectively.
    The last subplot shows the ability of the underlying ensemble model:
    the average IoU between vulnerable lines and those identified by the ensemble model, the proportions of missed vulnerable lines, i.e., false benign rate (FBR), and the proportions of misclassified benign lines, i.e., false vulnerable rate (FVR).
    }
    \label{fig:bleu_analysis}
\end{figure}

For convenient discussion, we set $\delta_{\textit{IoU}}$ to 0.5 in this experiment, based on the intuition from computer vision~\cite{Stewart:2016:EndToEndPeopleDetectioninCrowdedScenes}.
Figure~\ref{fig:bleu_analysis} shows the average \truevul's effectiveness across four models and four datasets at different $\delta_{\textit{BLEU}}$. 
We illustrate the changes in overall effectiveness in the first four subplots using Accuracy, AUC, F1-score, and G-mean, respectively.
The last subplot shows the ability of the underlying ensemble model.
We report the average IoU between vulnerable lines and those identified by the ensemble model, with false benign rate (FBR) and false vulnerable rate (FVR), the proportions of missed vulnerable lines and misclassified benign lines, respectively.

A higher $\delta_{\textit{BLEU}}$ retains more lines in the list of non-vulnerable lines, $L^-$, potentially introducing noise and making the ensemble model more likely to classify a line as a benign candidate.
As a result, Accuracy, Precision, Sensitivity, F1-score, and FBR increase, while Specificity and FVR decrease.
This indicates that a higher $\delta_{\textit{BLEU}}$ makes \truevul{} tend to misclassify trustworthy as untrustworthy.
Conversely, as $\delta_{\textit{BLEU}}$ decreases, Specificity and FVR increase, while Accuracy, Precision, Sensitivity, F1-score, and FBR decline.
Given a lower $\delta_{\textit{BLEU}}$, fewer lines are kept in $L^-$, making the ensemble model struggle to recognise benign candidates.
Hence, \truevul{} tends to misclassify predictions as trustworthy.
Figure~\ref{fig:bleu_analysis} shows that the optimal $\delta_{\textit{BLEU}}$ ranges from 0.4 to 0.6. 
In the subsequent experiments, we set $\delta_{\textit{BLEU}}$ to 0.5, as it balances the effectiveness across the different metrics.

\subsubsection{\textbf{Impact of IoU thresholds}}

\begin{figure}[t]
    \centering
    \includegraphics[scale=0.5, clip, trim=0mm 0mm 0 0]{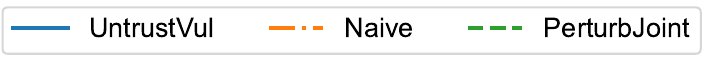}
    {\includegraphics[scale=0.5, clip, trim=0mm 0mm 0mm 0]{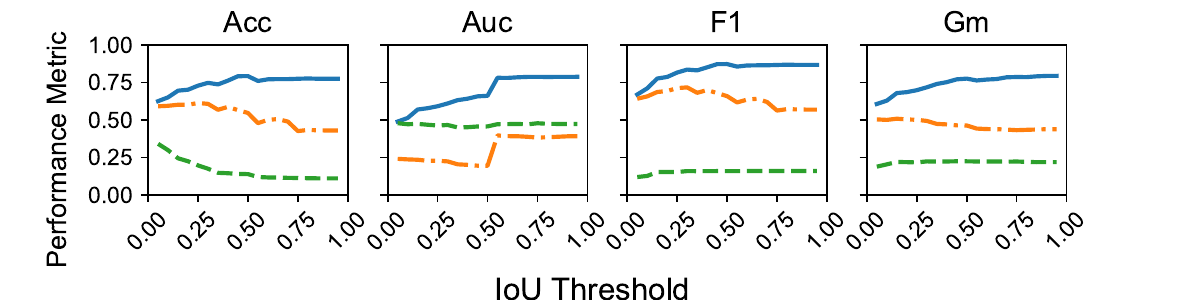}}
    \caption{Comparing the average effectiveness of \truevul{} and the baselines across different values of $\delta_{\textit{IoU}}$.}
    \label{fig:iou_analysis}
\end{figure}

Figure~\ref{fig:iou_analysis} presents the average effectiveness of \truevul{} and the baselines across four models and four datasets at various $\delta_{\textit{IoU}}$.
We compare against Naive and PerturbJoint, the most effective baseline among Rahman et al.'s approaches~\cite{Rahman:2024:TowardsCausalDeepLearningforVulnerabilityDetection}.
A lower $\delta_{\textit{IoU}}$ relaxes ground-truth trustworthiness labeling by allowing more irrelevant lines, resulting in more predictions labeled as trustworthy, while a higher $\delta_{\textit{IoU}}$ leads to more untrustworthy predictions.
We also report the changes in overall effectiveness using Accuracy, AUC, F1-score, and G-mean, respectively.

With \truevul, all metrics increase as $\delta_{\textit{IoU}}$ increases, showing better effectiveness under stricter trustworthiness labeling.
In contrast, the baselines exhibit more complex trends.
For Naive, AUC, Precision, and Specificity increase, while other metrics decrease.
For PerturbJoint, Accuracy drops, Precision increases, while the others show no significant changes.
Nevertheless, \truevul{} consistently outperforms the baselines across different $\delta_{\textit{IoU}}$.
In the subsequent experiments, we set $\delta_{\textit{IoU}}$ to 0.5, as it shows the optimal \truevul's effectiveness across the metrics.

\begin{custombox}[Takeaway~\#5]
Increasing $\delta_{\textit{BLEU}}$ introduces more noise into the training of the underlying ensemble model, while decreasing $\delta_{\textit{BLEU}}$ impairs its ability to identify benign candidates.
In addition, \truevul{} consistently outperforms all baselines under different values of $\delta_{\textit{IoU}}$.
\end{custombox}

\subsection{\textbf{RQ5: Improving Trustworthiness}}

We evaluate \truevul's ability to improve the trustworthiness of vulnerability detection models.
Follow CausalVul~\cite{Rahman:2024:TowardsCausalDeepLearningforVulnerabilityDetection}, we apply a causal learning algorithm for model retraining,
aiming to disable models from using spurious features, and encourage the use of correct features.
While CausalVul focuses on variable names and APIs, we inject irrelevant lines of code identified \truevul{} as dead code.

\begin{table}[t]
\centering
\setlength\tabcolsep{2.5pt} 
\caption{Comparing the effectiveness of \truevul{} (ours) and CausalVul~\cite{Rahman:2024:TowardsCausalDeepLearningforVulnerabilityDetection} in improving vulnerability detectors. 
Best results in each column are highlighted in grey}
\label{tab:improve}

\subfloat[CodeBERT]{\resizebox{\linewidth}{!}{%
\begin{tabular}{|l|ccc|ccc|ccc|ccc|}
\hline
          & \multicolumn{3}{c|}{BigVul} & \multicolumn{3}{c|}{MegaVul} & \multicolumn{3}{c|}{SARD} & \multicolumn{3}{c|}{PrimeVul} \\
          & F1     & $\overline{IoU}$    & $\overline{\mathcal{T}}$ & F1      & $\overline{IoU}$      & $\overline{\mathcal{T}}$     & F1     & $\overline{IoU}$    & $\overline{\mathcal{T}}$ & F1     & $\overline{IoU}$    & $\overline{\mathcal{T}}$    \\
\hline
Vanilla   & 0.38   & 0.21   & 0.42       & 0.28  &  0.12         &  0.37             &   0.09      &  0.04       &   \cellcolor{gray!30}{0.26}  & 0.27 & 0.15 &  0.39       \\
CausalVul~\cite{Rahman:2024:TowardsCausalDeepLearningforVulnerabilityDetection} & \cellcolor{gray!30}{0.41}   & 0.22     &  0.48        & 0.33        & 0.13       &   0.39           &  0.12      & 0.04       &  0.25    & 0.26 & 0.16 &  0.40       \\
\truevul   & \cellcolor{gray!30}{0.41}   & \cellcolor{gray!30}{0.27}   & \cellcolor{gray!30}{0.51}       &  \cellcolor{gray!30}{0.36}       &  \cellcolor{gray!30}{0.16}        &  \cellcolor{gray!30}{0.43}            & \cellcolor{gray!30}{0.15}       & \cellcolor{gray!30}{0.07}       & \cellcolor{gray!30}{0.26}   & \cellcolor{gray!30}{0.30} & \cellcolor{gray!30}{0.19} &  \cellcolor{gray!30}{0.40}         \\
\hline
\end{tabular}%
}}

\subfloat[GraphCodeBERT]{\resizebox{\linewidth}{!}{%
\begin{tabular}{|l|ccc|ccc|ccc|ccc|}
\hline
          & \multicolumn{3}{c|}{BigVul} & \multicolumn{3}{c|}{MegaVul} & \multicolumn{3}{c|}{SARD} & \multicolumn{3}{c|}{PrimeVul} \\
          & F1     & $\overline{IoU}$    & $\overline{\mathcal{T}}$ & F1      & $\overline{IoU}$      & $\overline{\mathcal{T}}$     & F1     & $\overline{IoU}$    & $\overline{\mathcal{T}}$ & F1     & $\overline{IoU}$    & $\overline{\mathcal{T}}$    \\
\hline
Vanilla & 0.38   & 0.21   & 0.48       & 0.32  &  0.12         &  0.40             &   0.02      &  0.02       &   \cellcolor{gray!30}{0.27}  & 0.29 & 0.08 &  0.32          \\
CausalVul~\cite{Rahman:2024:TowardsCausalDeepLearningforVulnerabilityDetection} & 0.39   & 0.23       &  0.49          &  0.32       & 0.12         & \cellcolor{gray!30}{0.42}             &  0.03       & 0.03       & 0.24      & {0.30} & 0.08 &  0.37       \\
\truevul & \cellcolor{gray!30}{0.40}      & \cellcolor{gray!30}{0.26}   & \cellcolor{gray!30}{0.50}       & \cellcolor{gray!30}{0.34}        &  0.12        &  \cellcolor{gray!30}{0.42}            & \cellcolor{gray!30}{0.04}       & \cellcolor{gray!30}{0.04}       & \cellcolor{gray!30}{0.27}     & \cellcolor{gray!30}{0.36} & \cellcolor{gray!30}{0.10} &  \cellcolor{gray!30}{0.38}        \\
\hline
\end{tabular}%
}}

\subfloat[UniXcoder]{\resizebox{\linewidth}{!}{%
\begin{tabular}{|l|ccc|ccc|ccc|ccc|}
\hline
          & \multicolumn{3}{c|}{BigVul} & \multicolumn{3}{c|}{MegaVul} & \multicolumn{3}{c|}{SARD} & \multicolumn{3}{c|}{PrimeVul} \\
          & F1     & $\overline{IoU}$    & $\overline{\mathcal{T}}$ & F1      & $\overline{IoU}$      & $\overline{\mathcal{T}}$     & F1     & $\overline{IoU}$    & $\overline{\mathcal{T}}$ & F1     & $\overline{IoU}$    & $\overline{\mathcal{T}}$    \\
\hline
Vanilla & 0.39   & 0.22   & 0.45       & 0.26  &  0.13         &  0.40             &   0.14      &  0.05       &   0.22     & 0.28 & 0.12 &  0.40       \\
CausalVul~\cite{Rahman:2024:TowardsCausalDeepLearningforVulnerabilityDetection} & 0.40   & 0.22       & 0.48           & \cellcolor{gray!30}{0.36}        & 0.14         &  0.49            &  0.56      &  0.06      & 0.23      & 0.29 & 0.15 &  0.42       \\
\truevul & \cellcolor{gray!30}{0.41}      &  \cellcolor{gray!30}{0.24}  & \cellcolor{gray!30}{0.52}       &  \cellcolor{gray!30}{0.36}       &  \cellcolor{gray!30}{0.16}        &  \cellcolor{gray!30}{0.50}            & \cellcolor{gray!30}{0.59}       &  \cellcolor{gray!30}{0.07}      &  \cellcolor{gray!30}{0.29}     & \cellcolor{gray!30}{0.38} & \cellcolor{gray!30}{0.15} &  \cellcolor{gray!30}{0.45}       \\
\hline
\end{tabular}%
}}
\vspace{-3mm}
\end{table}

For fair comparisons, we reuse the same experimental setup from CausalVul, using three SOTA models, CodeBERT~\cite{Feng:2020:CodeBERT:APreTrainedModelforProgrammingandNaturalLanguages}, GraphCodeBERT~\cite{Guo:2020:Graphcodebert:PreTrainingCodeRepresentationsWithDataFlow} and UniXcoder~\cite{Guo:2022:UniXcoder:UnifiedCrossModalPretrainingforCodeRepresentation}, as vanilla models. 
They are trained and evaluated on BigVul using the same data split, and further evaluated on MegaVul, SARD, and PrimeVul to assess generalisability.
Table~\ref{tab:improve} compares \truevul{} and CausalVul based on F1-score for the improved ability for vulnerability detection, as well as IoU and \truevul's $\mathcal{T}$ score for estimating trustworthiness.


\truevul{} improve vulnerability detection by 3\% in intra-project settings on BigVul.
Causal learning focuses on causal signals that compensate for the loss of spurious features, ultimately improving intra-project performance~\cite{Rahman:2024:TowardsCausalDeepLearningforVulnerabilityDetection}.
In cross-project settings on MegaVul and SARD, \truevul{} improves vulnerability detection by 45\%, showing the potential to improve generalisability. 
Since the spurious features may not be present in the cross-project data, learning to ignore them helps significantly improve performance.

\truevul{} enhances model trustworthiness by 6\% in $\overline{IoU}$ and 9\% in $\overline{\mathcal{T}}$.
Compared to CausalVul, \truevul{} generally performs better. 
While CausalVul occasionally attains equal F1-score, it often shows lower trustworthiness in $\overline{IoU}$ and $\overline{\mathcal{T}}$.
By ignoring spurious features beyond variable names and APIs, \truevul{} can significantly improve the trustworthiness, helping models focus on the root causes of the vulnerabilities instead of the spurious features.

\begin{figure}[t]
\subfloat[F1]{\includegraphics[scale=0.6, clip, trim=16.5mm 4mm 30mm 21mm]{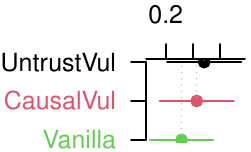}}
\hfill
\subfloat[$\overline{IoU}$]{\includegraphics[scale=0.6, clip, trim=16.5mm 4mm 30mm 21mm]{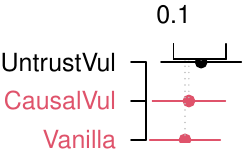}}
\hfill
\subfloat[$\overline{\mathcal{T}}$]{\includegraphics[scale=0.6, clip, trim=16.5mm 4mm 30mm 21mm]{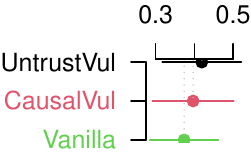}}
\caption{Comparing the ability to improve vulnerability detectors of \truevul{} and CausalVul~\cite{Rahman:2024:TowardsCausalDeepLearningforVulnerabilityDetection} using the Scott-Knott ESD test~\cite{Tantithamthavorn:2017:AnEmpiricalComparisonofModelValidationTechniquesforDefectPredictionModels}, with colors indicating statistically distinct clusters.}
\label{fig:ScottKnottESD_imp}
\end{figure}
We further perform the ESD test~\cite{Tantithamthavorn:2017:AnEmpiricalComparisonofModelValidationTechniquesforDefectPredictionModels} that partitions the approaches into statistically distinct groups with non-negligible difference. 
Figure~\ref{fig:ScottKnottESD_imp} shows that the models of \truevul{} are statistically significantly better than those of CausalVul and Vanilla in both performance (e.g., F1-score) and trustworthiness (e.g., $\overline{IoU}$ and $\overline{\mathcal{T}}$).
On the other hand, only improved F1-score by CausalVul is statistically significant compared to Vanilla, while the improved trustworthiness in $\overline{IoU}$ is not.

\begin{custombox}[Takeaway~\#6]
\truevul{} increases the ability to detect vulnerable code by up to 8\%, generalisation performance by up to 321\%, and trustworthiness by up to 100\%.
\end{custombox}

\section{Discussion}
\subsection{Detecting Untrustworthy Vulnerability Predictions}

This study confirms that SOTA vulnerability detection models often make untrustworthy predictions without fully reasoning about code~\cite{Ganz:2021:ExplainingGraphNeuralNetworksforVulnerabilityDiscovery, Chakraborty:2022:DeepLearningBasedVulnerabilityDetectionAreWeThereYet, Steenhoek:2023:AnEmpiricalStudyofDeepLearningModelsforVulnerabilityDetection, Cheng:2024:BeyondFidelityExplainingVulnerabilityLocalizationofLearningBasedDetectors}.
Such predictions can mislead developers by highlighting irrelevant lines, leading to ineffective or even harmful patches.
Hence, despite high performance, it is also important to assess their decision-making.
We introduce \truevul{}, the first automated approach for detecting untrustworthy vulnerability predictions.
\truevul's effectiveness
indicates that analysing the syntax and dependencies can reveal untrustworthy behavior.
Automated trustworthiness assessment is an emerging yet essential challenge, not only for vulnerability detection but also for other ML tasks.
Yet, this problem remains underexplored and demands further attention.

{\color{revisedcolor}\subsection{Limitations}
\truevul{} sometimes fails to distinguish between syntactically similar benign and vulnerable lines.
In such cases, a prediction may highlight a benign line that closely resembles a vulnerable one.
This form of untrustworthiness is subtle and challenging even for human auditors.
As a result, \truevul{} may mistakenly consider the prediction trustworthy, despite the vulnerability located at a different line.

The main reason is that \truevul's first stage, line-level assessment, can overestimate an isolated benign line as vulnerable if it looks like one.
Several strategies can be explored in the future to mitigate this limitation.
For example, incorporating semantic encoding that captures surrounding context can advance \truevul's code understanding.
Additional static analyses, such as taint analysis~\cite{Liu:2025:LLMPoweredStaticBinaryTaintAnalysis} or symbolic execution~\cite{Baldoni:2018:ASurveyofSymbolicExecutionTechniques}, as well as dynamic analysis~\cite{Wang:2024:CombiningStructuredStaticCodeInformationandDynamicSymbolicTracesforSoftwareVulnerabilityPrediction}, may further help differentiate benign from vulnerable lines.
We believe these improvements benefit not only \truevul{} but also broader vulnerability analysis research.
}

\subsection{Implications to Software Engineering}

ML has been widely applied in SE to automate many tasks in the lifecycle, such as planning, design, coding, and testing.
However, growing concerns around trustworthiness hinder its broader adoption~\cite{Eoychoudhury:2025:AISoftwareEngineerProgrammingwithTrust}.
While making ML-based systems fully trustworthy is the ultimate goal, achieving this remains challenging.
Hence, we believe a practical approach to enhance their real-world usability is to provide alerts when predictions are untrustworthy.

Integrating \truevul{} into the SE lifecycle can significantly enhance the development and deployment of ML vulnerability detectors. 
\truevul{} can serve as an automated oracle for trustworthiness testing~\cite{Lam:2024:AutomatedTrustworthinessOracleGenerationforMachineLearningTextClassifiers}.
This testing process is crucial and closely intertwined with other SE lifecycle activities~\cite{Riccio:2020:TestingMachineLearningBasedSystemsASystematicMapping}, particularly in the iterative development of ML-based systems, where performance and trustworthiness must be evaluated and refined~\cite{Silverio:2022:SoftwareEngineeringforAIBasedSystemsASurvey}. 
By early detecting trustworthiness issues during development, it helps prevent deployment failures.
It also supports feedback-driven repair, as demonstrated in Section~\ref{sec:expr_results}, reducing human effort.

{\color{revisedcolor}
\section{Threats to Validity}
This section addresses the potential threats and outlines the mitigation strategies we employed to minimize their impact.

\subsection{Threats to Internal Validity} \label{sec:Threats2Internal}

\subsubsection{Reliance on Joern}
We used Joern to parse each function into a PDG in the dependency-level assessment.
Although Joern is an SOTA static analysis tool with high accuracy and has been widely used for vulnerability-related tasks~\cite{Cheng:2024:BeyondFidelityExplainingVulnerabilityLocalizationofLearningBasedDetectors}, it still has limitations.
In particular, it struggles with certain C/C++ constructs, such as complex pointer arithmetic, overloaded constructors, and indirect calls.
Consequently, some dependencies may be omitted or misrepresented, which affects \truevul's analysis.
To mitigate this, we discard functions for which Joern fails to generate PDGs across the datasets.
Nevertheless, we claim that future empirical investigations may provide further insights into the robustness and completeness of Joern-generated PDGs.
They may also introduce better tools that can be integrated into \truevul.

\subsubsection{Noise in ground-truth line-level labels}
In SARD, vulnerable lines are explicitly annotated.
In BigVul, MegaVul, and PrimeVul, while such explicit annotations are unavailable, they can be inferred from vulnerability-fixing commits. 
Following prior research~\cite{Chu:2024:GraphNeuralNetworksforVulnerabilityDetectionACounterfactualExplanation, Fan:2020:BigVul:ACCCodeVulnerabilityDatasetwithCodeChangesandCVESummaries, 
Hin:2022:Linevd:StatementLevelVulnerabilityDetectionUsingGraphNeuralNetworks, 
Hu:2023:InterpretersforGNNBasedVulnerabilityDetectionAreWeThereYet, 
Li:2021:VulnerabilityDetectionwithFineGrainedInterpretations}, we treat lines deleted or modified in these commits, and those control- or data-dependent on added lines, as vulnerable.
Although this practice is widely adopted and empirically supported in the prior studies, some lines may not actually be vulnerable.
This introduces noise into the ground truths and affects the evaluation.
To mitigate this, we applied multiple filtering rules~\cite{Hin:2022:Linevd:StatementLevelVulnerabilityDetectionUsingGraphNeuralNetworks} to exclude redundant lines, such as comments, blank lines, delimiters, and keyword-only.
Future work may explore more accurate approaches to obtain higher-quality line-level ground truths.


\subsubsection{Potential bias in line-level assessment}
This threat stems from the historical data and transformer-based encoders.

The \textit{historical data} is derived from BigVul, and we applied several cleaning steps to mitigate bias.
First, we mitigated tangled commits by following the approach in~\cite{Wang:2024:ReposVul:ARepositoryLevelHighQualityVulnerabilityDataset}.
We provided an LLM with the function, code changes, and supplementary information from the CVE database.
We then prompted it to evaluate the relevance between the code changes and the vulnerability based on explicit mentions in the supplementary information~\cite{Ding:2024:VulnerabilityDetectionwithCodeLanguageModelsHowFarAreWe, Wang:2024:ReposVul:ARepositoryLevelHighQualityVulnerabilityDataset}. 
Since human experts have analyzed the CVE records, the supplementary information is a reliable reference.
Unrelated changes were further filtered using four static analysis tools
and 
multiple heuristics~\cite{Hin:2022:Linevd:StatementLevelVulnerabilityDetectionUsingGraphNeuralNetworks}.
We also applied a BLEU threshold to remove conflict and duplicate samples between the list of vulnerable and non-vulnerable lines.
The sensitivity of this threshold is analysed in Section~\ref{sec:sensitivity_analysis}.  
These procedures follow prior studies and have been validated~\cite{Ding:2024:VulnerabilityDetectionwithCodeLanguageModelsHowFarAreWe, Wang:2024:ReposVul:ARepositoryLevelHighQualityVulnerabilityDataset, Hin:2022:Linevd:StatementLevelVulnerabilityDetectionUsingGraphNeuralNetworks}.
However, we acknowledge that the LLM may hallucinate, and static analysis tools are imperfect.
Consequently, some changes unrelated to vulnerabilities may be mislabeled as historical vulnerable lines.
This leaves a small portion of mislabeled historical data.
However, we believe all the reported results will still hold good.

The \textit{transformer-based encoders} for line-level classification may inherit biases from their pretraining corpora. 
We mitigate this risk through ensemble learning~\cite{Leon:2017:EvaluatingTheEffectOfVotingMethodsOnEnsemble-basedClassification}, aggregating multiple SOTA pre-trained code models~\cite{Feng:2020:CodeBERT:APreTrainedModelforProgrammingandNaturalLanguages, Guo:2022:UniXcoder:UnifiedCrossModalPretrainingforCodeRepresentation, Guo:2020:Graphcodebert:PreTrainingCodeRepresentationsWithDataFlow}.

\subsection{Threats to External Validity}

\subsubsection{Language generalisability}
Our core methodology is inherently language-agnostic.
Joern already supports multiple languages, such as C/C++, Java, Python, and JavaScript.
As most existing vulnerability detection research~\cite{Ding:2024:VulnerabilityDetectionwithCodeLanguageModelsHowFarAreWe} focuses on C/C++, we also centre our evaluation on C/C++. 
Adapting \truevul{} to other languages would only require curating language-specific historical data for line-level assessment. 
Future work could explore the effectiveness and generalisability of \truevul{} across different programming languages.

\subsubsection{Vulnerability generalisability}
The second threat to external validity is 
how well \truevul{} handles a wide range of vulnerabilities.
We addressed this by evaluating four diverse datasets, BigVul, MegaVul, SARD, and PrimeVul, which together contain over 80K vulnerabilities spanning more than 140 CWEs, including both synthetic and real-world vulnerabilities from 2002 to 2023. 
Future work can examine more recent vulnerabilities to further assess \truevul.

\subsubsection{Model generalisability}
\truevul’s effectiveness may vary across different ML-based vulnerability detectors.
To mitigate this, we evaluated it on both SOTA transformer- and graph-based models over 115K predictions.
We also employed \truevul{} with various interpretation methods, such as attention and GNNExplainer, to further assess its robustness.
We acknowledge that evaluating additional models is beneficial to more comprehensively assess \truevul.

\subsubsection{Industrial-scale applicability}
On three real-world datasets, BigVul, MegaVul, and PrimeVul, \truevul{} takes 1.35--1.52 seconds to process a function with an average size of 119-167 LOC. 
These results indicate the practical efficiency of \truevul. 
Given the average size of 350,542 LOC of open-source software reported in prior work~\cite{Woo:2021:CENTRIS:APreciseandScalableApproachforIdentifyingModifiedOpenSourceSoftwareReuse}, scanning an entire project would take approximately 3,190 seconds.
This overhead is acceptable, as it is outweighed by the benefits of fostering trust in security-critical tasks.
This demonstrates the feasibility and applicability of \truevul{} in real-world settings.
Integrating \truevul{} into industrial pipelines in future work would further validate its practical usefulness.

\subsection{Threats to Construct Validity}

Threats to {construct validity} can arise from the ground truths used to evaluate \truevul{} that are based on \textit{Intersection over Union} (IoU).
We acknowledge that IoU does not directly measure trustworthiness, but approximates it through the overlap between lines in model interpretations and ground-truth vulnerable lines.
It indicates how well the interpretations correctly highlight vulnerable lines.
Although IoU is a well-understood, standard evaluation metric for object detection tasks, including assessing finer-grained vulnerability detection~\cite{Jiang:2025:EnhancingFineGrainedVulnerabilityDetectionWithReinforcementLearning}, it is sensitive to noise and annotation style, which may limit its ability to capture trustworthiness.
To mitigate this, we reduced noise in ground-truth vulnerable lines through multiple cleaning steps as described in Section~\ref{sec:Threats2Internal}.
We also conducted a sensitivity analysis of the IoU threshold on \truevul's effectiveness in Section~\ref{sec:sensitivity_analysis}.
Future work can include a comprehensive manual review to assess how well IoU approximates trustworthiness.
}

\section{Related Work}
\label{sec:related_work}
\subsection{ML-based Vulnerability Detection Models}

ML vulnerability detectors can be categorised as follows. 

\subsubsection{Token-based models}
These models, such as VulDeePecker~\cite{Li:2018:VulDeePecker:ADeepLearningBasedSystemforVulnerabilityDetection} and SySeVR~\cite{Li:2022:SySeVR:AFrameworkforUsingDeepLearningtoDetectSoftwareVulnerabilities}, treat code as a sequence of tokens, embedding them as vectors for training using architectures like BiLSTM.
Recent models, such as LineVul~\cite{Fu:2022:LineVul:ATransformerBasedLineLevelVulnerabilityPrediction} and SVulD~\cite{Ni:2023:DistinguishingLookAlikeInnocentandVulnerableCodebySubtleSemanticRepresentationLearningandExplanation}, leverage pre-trained transformers, e.g., CodeBERT~\cite{Feng:2020:CodeBERT:APreTrainedModelforProgrammingandNaturalLanguages}, 
GraphCodeBERT~\cite{Guo:2020:Graphcodebert:PreTrainingCodeRepresentationsWithDataFlow},
and UniXcoder~\cite{Guo:2022:UniXcoder:UnifiedCrossModalPretrainingforCodeRepresentation}, 
for improved performance.

\subsubsection{Graph-based models}
These models represent code as graphs that capture abstract syntax trees and dependencies, enabling training on both semantic and syntactic information~\cite{Chu:2024:GraphNeuralNetworksforVulnerabilityDetectionACounterfactualExplanation}.
For instance, ReVeal~\cite{Chakraborty:2022:DeepLearningBasedVulnerabilityDetectionAreWeThereYet} and IVDetect~\cite{Li:2021:VulnerabilityDetectionwithFineGrainedInterpretations}
leverage GNNs and achieve accuracy of up to 90\%. 

Despite the promising results of ML-based vulnerability detectors, existing end-to-end training approaches primarily focus on performance in held-out evaluations and do not ensure that predictions are made based on the correct indicators.
Recent research has therefore begun investigating the reasoning behind these predictions.
Our work, \truevul{}, addresses this gap by automatically assessing and detecting untrustworthy predictions, helping to avoid overestimating the quality of ML-based vulnerability detectors.

\subsection{Interpretability of ML-based Vulnerability Detectors}

Due to the lack of explainability, methods have been introduced to interpret predictions by identifying important input features.
Some use gradients~\cite{Simonyan:2014:DeepInsideConvolutionalNetworks:VisualisingImageClassificationModelsandSaliencyMaps, Selvaraju:2017:GradCAM:VisualExplanationsFromDeepNetworksviaGradientBasedLocalization} or the attention mechanism~\cite{Hin:2022:Linevd:StatementLevelVulnerabilityDetectionUsingGraphNeuralNetworks} to approximate the feature importance. 
Other methods~\cite{Li:2017:UnderstandingNeuralNetworksthroughRepresentationErasure, Ying:2019:GNNExplainer:GeneratingExplanationsforGraphNeuralNetworks} perturb inputs and evaluate the model output changes.
Recent methods~\cite{Ribeiro:2016:WhyShouldITrustYouExplainingthePredictionsofAnyClassifier, Huang:2023:GraphLIME:LocalInterpretableModelExplanationsforGraphNeuralNetworks} locally approximate the model as an interpretable surrogate model.
These methods have been used to locate vulnerabilities, e.g.,
LineVul~\cite{Fu:2022:LineVul:ATransformerBasedLineLevelVulnerabilityPrediction} and mVulPreter~\cite{Zou:2022:mVulPreter:AMultiGranularityVulnerabilityDetectionSystemWithInterpretations} leverage self-attention,
while IVDetect~\cite{Li:2021:VulnerabilityDetectionwithFineGrainedInterpretations} uses GNNExplainer~\cite{Ying:2019:GNNExplainer:GeneratingExplanationsforGraphNeuralNetworks}.

\lam{\st{Importantly, these interpretation methods differ from UntrustVul. Instead of focusing on generating interpretations for predictions, UntrustVul automatically evaluates the plausibility and reasonableness of the generated interpretations. These two are complementary rather than interchangeable. UntrustVul can work with any interpretation method, each providing different perspectives on the same prediction.}}

\ehl{
Existing interpretability research primarily aims to generate \textit{faithful} interpretations that correctly describe the model reasoning~\cite{Ying:2019:GNNExplainer:GeneratingExplanationsforGraphNeuralNetworks, Ganz:2021:ExplainingGraphNeuralNetworksforVulnerabilityDiscovery, Chu:2024:GraphNeuralNetworksforVulnerabilityDetectionACounterfactualExplanation}.
In contrast, our focus is on trustworthiness through the lens of \textit{plausibility}, i.e., how reasonable the interpretations are.
In vulnerable detection, it means the prediction highlights lines of code relevant to the actual vulnerability. 
There may be a limited number of studies~\cite{Lam:2024:AutomatedTrustworthinessOracleGenerationforMachineLearningTextClassifiers} in other domains that automatically assess plausibility or prediction trustworthiness. 
However, they do not apply to vulnerability detectors. 
To our knowledge, no existing method can automatically evaluate plausibility in vulnerability detection.
}

\ehl{
Crucially, \truevul{} does not compete with these interpretation methods. 
Rather, it addresses a complementary problem. 
Instead of producing interpretations, \truevul{} evaluates whether the interpretations generated by any interpretation method are plausible and reflect meaningful reasoning. 
In this way, \truevul{} acts as an independent trustworthiness check that can be paired with any interpretation method, such as attention or GNNExplainer. 
Each provides a different lens on the model’s behaviour, and \truevul{} helps assess the trustworthiness of those lenses.
}

\subsection{Trustworthiness of ML-based Vulnerability Detectors}

\subsubsection{Detecting trustworthiness issues}

In recent years, the trustworthiness problem in ML has been explored in various areas~\cite{Kästner:2021:OntheRelationofTrustandExplainabilityWhytoEngineerforTrustworthiness, Lam:2024:AutomatedTrustworthinessOracleGenerationforMachineLearningTextClassifiers}. 
Despite this progress, only a few methods exist for detecting trustworthiness issues, most of which still rely on human validation~\cite{Lapuschkin:2019:UnmaskingCleverHansPredictorsAndAssessingWhatMachinesReallyLearn, Schramowski:2020:MakingDeepNeuralNetworksRightForTheRightScientificReasonsByInteractingWithTheirExplanations, Ghai:2021:ExplainableActiveLearningXALTowardAIExplanationsasInterfacesforMachineTeachers}.

In the context of vulnerability detection, Tien et al.~\cite{Tien:2022:HumanInTheLoopXAIEnabledVulnerabilityDetectionInvestigationAndMitigation} introduced 
a conceptual framework that visualises model explanations for human experts to verify in each iteration, and provides a way to adjust the input and parameters in the next iteration.
Several studies~\cite{Ganz:2021:ExplainingGraphNeuralNetworksforVulnerabilityDiscovery, Chakraborty:2022:DeepLearningBasedVulnerabilityDetectionAreWeThereYet, Steenhoek:2023:AnEmpiricalStudyofDeepLearningModelsforVulnerabilityDetection, Cheng:2024:BeyondFidelityExplainingVulnerabilityLocalizationofLearningBasedDetectors} conducted manual examinations of explanations of vulnerability predictions to understand what the models learn.
They have reported that these models often rely on spurious correlations rather than root causes for predictions.
These spurious correlations come from tokens frequently appearing in training data, such as identifiers, keywords, and delimiters.

While recent studies aim to detect trustworthiness issues in vulnerability predictions automatically, they typically operate under specific assumptions.
For example, identifiers, such as variable names and function names, are viewed as potential sources of spurious correlations with the ``vulnerable'' class, resulting in untrustworthy predictions~\cite{Gao:2023:TwoSidesoftheSameCoinExploitingtheImpactofIdentifiersinNeuralCodeComprehension, Rahman:2024:TowardsCausalDeepLearningforVulnerabilityDetection}. 
In contrast, \truevul{} automatically detects trustworthiness issues beyond the identifier level, identifying them at both the statement and dependency levels.
\truevul{} assesses not only whether prediction-contributing lines truly align with patterns linked to vulnerabilities, but also whether these statements exhibit dependencies critical to the predicted vulnerability.

\subsubsection{Improve the trustworthiness}

Multiple approaches have been introduced to counteract spurious correlations.
For example, Gao et al.~\cite{Gao:2023:TwoSidesoftheSameCoinExploitingtheImpactofIdentifiersinNeuralCodeComprehension} mitigated the misleading information from identifiers via counterfactual inference.
Imgrund et al.~\cite{Imgrund:2023:BrokenPromisesMeasuringConfoundingEffectsinLearningbasedVulnerabilityDiscovery} proposed normalising code with a consistent code style, tokenising inputs, and applying causal learning.
Rahman et al.~\cite{Rahman:2024:TowardsCausalDeepLearningforVulnerabilityDetection} designed perturbations to reveal overreliance on identifiers and applied causal learning to mitigate it. 

While existing studies rely on manual identification or certain assumptions about spurious correlations, \truevul{} can automatically identify untrustworthy predictions and improve model trustworthiness without human intervention.

\section{Conclusion and Future Work}
\ehl{We propose \truevul, the first automated approach to detect untrustworthy vulnerability predictions by leveraging supervised learning, static analysis, and rule-based analysis.
Experiments on multiple SOTA vulnerability detectors and datasets show \truevul's strong performance in detecting untrustworthy predictions and improving trustworthiness.
\truevul{} also offers multiple practical benefits for software security workflows.
For example, filtering out untrustworthy predictions saves developers from chasing irrelevant lines of code, reduces time wasted on misleading interpretations, and helps developers avoid mispatches.
}

\ehl{While still in its early stages, \truevul{} introduces considerable opportunities for further exploration.
First, we can address label noise to improve the performance of \truevul.
Second, while the approach itself is general, our implementation and evaluation currently focus on C/C++ vulnerabilities.
Hence, generalising \truevul{} to other programming languages is an important avenue.
Third, we can integrate \truevul{} into IDEs and CI pipelines to support continuous and trustworthy vulnerability assessment. 
Finally, extending \truevul{} to other SE tasks can further foster human trust in ML4SE systems.
}


\bibliographystyle{IEEEtran}
\bibliography{sample-base}


 






\appendix[Supplementary Material] 
\label{sec:Supplementary_Material}
\subsection{Detailed Results of Effectiveness}

Tables~\ref{table:full_expr_results}--\ref{table:full_expr_results_cp} compare the effectiveness of \truevul{} and the baselines in intra-project and cross-project settings, i.e., on BigVul and on MegaVul, SARD, PrimeVul, respectively.
\begin{table*}
\centering
\caption{The \textit{intra-project} effectiveness of \truevul{} (ours) and the baselines on BigVul dataset}
\vspace{-3mm}
\label{table:full_expr_results}
\setlength\tabcolsep{0pt} 

\subfloat[LineVul]{\resizebox{0.312\linewidth}{!}{%
\begin{tabular}{|@{\hspace{2pt}}p{1.8cm}|CCCCCCC|}
\hline
             & Acc  & Auc   & Pre  & Sen  & F1   & Spe  & Gm   \\
\hline
\truevul & 0.78 & 0.86  & 0.97 & 0.77 & 0.86 & 0.83 & 0.80 \\
Naive        & 0.46 & 0.32  & 0.82 & 0.48 & 0.60 & 0.32 & 0.39 \\
PerturbVar   & 0.12 & 0.27  & 0.31 & 0.01 & 0.03 & 0.82 & 0.10 \\
PerturbAPI   & 0.14 & 0.46  & 0.59 & 0.01 & 0.02 & 0.95 & 0.10 \\
PerturbJoint & 0.14 & 0.42  & 0.57 & 0.01 & 0.02 & 0.95 & 0.10 \\
\hline
\end{tabular}%
}}
\hfill
\subfloat[SVulD]{\resizebox{0.225\linewidth}{!}{%
\begin{tabular}{|CCCCCCC|}
\hline
Acc  & Auc   & Pre  & Sen  & F1   & Spe  & Gm   \\
\hline
0.74 & 0.77  & 0.96 & 0.71 & 0.82 & 0.89 & 0.79 \\
0.58 & 0.60  & 0.85 & 0.58 & 0.69 & 0.57 & 0.58 \\
0.44 & 0.56  & 0.84 & 0.38 & 0.52 & 0.70 & 0.51 \\
0.40 & 0.55  & 0.83 & 0.33 & 0.47 & 0.72 & 0.48 \\
0.45 & 0.56  & 0.84 & 0.40 & 0.54 & 0.68 & 0.52 \\
\hline
\end{tabular}%
}}
\hfill
\subfloat[IVDetect]{\resizebox{0.225\linewidth}{!}{%
\begin{tabular}{|CCCCCCC|}
\hline
Acc  & Auc & Pre  & Sen  & F1   & Spe & Gm   \\
\hline
0.84 & 0.81  & 0.99 & 0.85 & 0.91 & 0.72 & 0.78 \\
0.32 & 0.32  & 0.94 & 0.31 & 0.47 & 0.48 & 0.38 \\
0.05 & 0.31  & 0.84 & 0.02 & 0.04 & 0.91 & 0.13 \\
0.05 & 0.35  & 0.89 & 0.01 & 0.02 & 0.97 & 0.10 \\
0.06 & 0.32  & 0.86 & 0.02 & 0.04 & 0.91 & 0.14 \\
\hline
\end{tabular}%
}}
\hfill
\subfloat[ReVeal]{\resizebox{0.225\linewidth}{!}{%
\begin{tabular}{|CCCCCCC|}
\hline
Acc  & Auc   & Pre  & Sen  & F1   & Spe  & Gm   \\
\hline
0.84 & 0.83  & 0.99 & 0.85 & 0.91 & 0.74 & 0.79 \\
0.36 & 0.31  & 0.93 & 0.35 & 0.51 & 0.40 & 0.38 \\
0.05 & 0.28  & 0.87 & 0.01 & 0.02 & 0.96 & 0.11 \\
0.05 & 0.32  & 1.00 & 0.01 & 0.01 & 1.00 & 0.07 \\
0.05 & 0.30  & 0.90 & 0.01 & 0.02 & 0.97 & 0.10 \\
\hline
\end{tabular}%
}}
\end{table*}

\begin{table*}
\centering
\caption{The \textit{cross-project} effectiveness of \truevul{} (ours) and the baselines on MegaVul, SARD, and PrimeVul}
\label{table:full_expr_results_cp}
\vspace{-3mm}
\setlength\tabcolsep{0pt} 

\subfloat[LineVul - MegaVul]{\resizebox{0.312\linewidth}{!}{%
\begin{tabular}{|@{\hspace{2pt}}p{1.8cm}|CCCCCCC|}
\hline
             & Acc  & Auc   & Pre  & Sen  & F1   & Spe  & Gm   \\
\hline
\truevul & 0.83 & 0.88 & 0.99 & 0.84 & 0.90 & 0.76 & 0.80 \\
Naive        & 0.25 & 0.34 & 0.96 & 0.25 & 0.39 & 0.56 & 0.37 \\
PerturbVar   & 0.14 & 0.32 & 0.95 & 0.13 & 0.23 & 0.68 & 0.30 \\
PerturbAPI   & 0.04 & 0.67 & 0.83 & 0.03 & 0.05 & 0.72 & 0.14 \\
PerturbJoint & 0.13 & 0.67 & 0.95 & 0.12 & 0.21 & 0.72 & 0.29 \\
\hline
\end{tabular}%
}}
\hfill
\subfloat[SVulD - MegaVul]{\resizebox{0.225\linewidth}{!}{%
\begin{tabular}{|CCCCCCC|}
\hline
Acc  & Auc   & Pre  & Sen  & F1   & Spe  & Gm   \\
\hline
0.82 & 0.82 & 0.98 & 0.83 & 0.89 & 0.75 & 0.79 \\
0.47 & 0.46 & 0.92 & 0.47 & 0.62 & 0.49 & 0.48 \\
0.33 & 0.52 & 0.97 & 0.32 & 0.48 & 0.67 & 0.46 \\
0.29 & 0.51 & 0.97 & 0.27 & 0.43 & 0.73 & 0.45 \\
0.34 & 0.51 & 0.97 & 0.33 & 0.49 & 0.67 & 0.47 \\
\hline
\end{tabular}%
}}
\hfill
\subfloat[IVDetect - MegaVul]{\resizebox{0.225\linewidth}{!}{%
\begin{tabular}{|CCCCCCC|}
\hline
Acc  & Auc & Pre  & Sen  & F1   & Spe & Gm   \\
\hline
0.81 & 0.74 & 0.99 & 0.81 & 0.89 & 0.69 & 0.75 \\
0.22 & 0.30 & 0.97 & 0.21 & 0.35 & 0.59 & 0.36 \\
0.03 & 0.30 & 0.93 & 0.01 & 0.02 & 0.96 & 0.09 \\
0.02 & 0.32 & 1.00 & 0.01 & 0.01 & 1.00 & 0.08 \\
0.03 & 0.32 & 0.97 & 0.02 & 0.03 & 0.98 & 0.12 \\
\hline
\end{tabular}%
}}
\hfill
\subfloat[ReVeal - MegaVul]{\resizebox{0.225\linewidth}{!}{%
\begin{tabular}{|CCCCCCC|}
\hline
Acc  & Auc   & Pre  & Sen  & F1   & Spe  & Gm   \\
\hline
0.72 & 0.70 & 0.99 & 0.72 & 0.84 & 0.63 & 0.68 \\
0.37 & 0.36 & 0.97 & 0.37 & 0.54 & 0.46 & 0.41 \\
0.02 & 0.32 & 0.88 & 0.01 & 0.01 & 0.99 & 0.09 \\
0.02 & 0.37 & 1.00 & 0.01 & 0.01 & 1.00 & 0.08 \\
0.03 & 0.35 & 0.96 & 0.01 & 0.01 & 0.99 & 0.10 \\
\hline
\end{tabular}%
}}


\subfloat[LineVul - SARD]{\resizebox{0.312\linewidth}{!}{%
\begin{tabular}{|@{\hspace{2pt}}p{1.8cm}|CCCCCCC|}
\hline
             & Acc  & Auc   & Pre  & Sen  & F1   & Spe  & Gm   \\
\hline
\truevul & 0.84 & 0.84    & 1.00    &  0.83   & 0.91   & 1.00    & 0.91 \\
Naive        & 0.80  & 0.76     & 0.97    & 0.81     & 0.88    & 0.75    & 0.78  \\
PerturbVar   & 0.36  &  0.78   & 1.00    & 0.29    & 0.45   & 1.00    & 0.54  \\
PerturbAPI   & 0.24 & 0.91    & 1.00    & 0.17    & 0.29   & 1.00    & 0.42 \\
PerturbJoint & 0.27 & 0.86    & 1.00    & 0.20   & 0.33   & 1.00    & 0.44  \\
\hline
\end{tabular}%
}}
\hfill
\subfloat[SVulD - SARD]{\resizebox{0.225\linewidth}{!}{%
\begin{tabular}{|CCCCCCC|}
\hline
Acc  & Auc   & Pre  & Sen  & F1   & Spe  & Gm   \\
\hline
0.76    & 0.82    & 1.00     &  0.65   & 0.78   & 0.94    & 0.78   \\
0.63    & 0.56    & 0.97    & 0.64     & 0.77    & 0.47    & 0.55   \\
0.13    & 0.56    & 0.97    & 0.10    & 0.18   & 0.90    &  0.3   \\
0.12    & 0.55    & 0.96    & 0.09    & 0.16   & 0.91    & 0.29   \\
0.14    & 0.55    & 0.96    & 0.12    & 0.21    & 0.88    & 0.32 \\
\hline
\end{tabular}%
}}
\hfill
\subfloat[IVDetect - SARD]{\resizebox{0.225\linewidth}{!}{%
\begin{tabular}{|CCCCCCC|}
\hline
Acc  & Auc & Pre  & Sen  & F1   & Spe & Gm   \\
\hline
0.85 & 0.82 & 0.99 & 0.86 & 0.92 & 0.68 & 0.76 \\
0.19 & 0.23 & 0.92 & 0.18 & 0.30 & 0.55 & 0.31 \\
0.04 & 0.23 & 0.77 & 0.01 & 0.02 & 0.90 & 0.10 \\
0.03 & 0.23 & 0.25 & 0.00 & 0.00 & 0.98 & 0.00 \\
0.04 & 0.24 & 0.76 & 0.01 & 0.02 & 0.92 & 0.09 \\
\hline
\end{tabular}%
}}
\hfill
\subfloat[ReVeal - SARD]{\resizebox{0.225\linewidth}{!}{%
\begin{tabular}{|CCCCCCC|}
\hline
Acc  & Auc   & Pre  & Sen  & F1   & Spe  & Gm   \\
\hline
0.75 & 0.71 & 0.99 & 0.75 & 0.85 & 0.61 & 0.68 \\
0.23 & 0.28 & 0.94 & 0.22 & 0.35 & 0.50 & 0.33 \\
0.03     & 0.28     & 0.80     & 0.01     & 0.01     & 0.99     & 0.03     \\
0.03     & 0.29     & 1.00     & 0.00     & 0.00     & 1.00     & 0.00     \\
0.03     & 0.29     & 1.00     & 0.01     & 0.01     & 1.00     & 0.03     \\
\hline
\end{tabular}%
}}

\subfloat[LineVul - PrimeVul]{\resizebox{0.312\linewidth}{!}{%
\begin{tabular}{|@{\hspace{2pt}}p{1.8cm}|CCCCCCC|}
\hline
             & Acc  & Auc   & Pre  & Sen  & F1   & Spe  & Gm   \\
\hline
\truevul & 0.85 & 0.91    & 0.99    &  0.84   & 0.91   & 0.93    & 0.88 \\
Naive        & 0.53    & 0.57    & 0.91    & 0.53     & 0.66    & 0.56    & 0.54   \\
PerturbVar   & 0.19  &  0.55   & 0.92    & 0.10    & 0.18   & 0.93    & 0.30  \\
PerturbAPI   & 0.12 & 0.75    & 0.81    & 0.01    & 0.03   & 0.97    & 0.12 \\
PerturbJoint & 0.18 & 0.76    & 0.96    & 0.09   & 0.16   & 0.97    & 0.29  \\
\hline
\end{tabular}%
}}
\hfill
\subfloat[SVulD - PrimeVul]{\resizebox{0.225\linewidth}{!}{%
\begin{tabular}{|CCCCCCC|}
\hline
Acc  & Auc   & Pre  & Sen  & F1   & Spe  & Gm   \\
\hline
0.83    & 0.88    & 0.99     &  0.83   & 0.90   & 0.88    & 0.86   \\
0.60    & 0.60    & 0.95    & 0.61     & 0.74    & 0.53    & 0.57   \\
0.25  &  0.57   & 0.95    & 0.21    & 0.34   & 0.85    & 0.42  \\
0.20 & 0.56    & 0.93    & 0.15    & 0.26   & 0.86    & 0.36 \\
0.24 & 0.57    & 0.94    & 0.20   & 0.33   & 0.82    & 0.40  \\
\hline
\end{tabular}%
}}
\hfill
\subfloat[IVDetect - PrimeVul]{\resizebox{0.225\linewidth}{!}{%
\begin{tabular}{|CCCCCCC|}
\hline
Acc  & Auc & Pre  & Sen  & F1   & Spe & Gm   \\
\hline
0.79 & 0.82 & 0.99 & 0.79 & 0.88 & 0.80 & 0.79 \\
0.50 & 0.31 & 0.97 & 0.51 & 0.67 & 0.27 & 0.37 \\
0.03 & 0.25 & 1.00 & 0.01 & 0.02 & 1.00 & 0.09 \\
0.03 & 0.32 & 0.67 & 0.00 & 0.01 & 0.96 & 0.00 \\
0.04 & 0.27 & 0.92 & 0.02 & 0.03 & 0.96 & 0.12 \\
\hline
\end{tabular}%
}}
\hfill
\subfloat[ReVeal - PrimeVul]{\resizebox{0.225\linewidth}{!}{%
\begin{tabular}{|CCCCCCC|}
\hline
Acc  & Auc   & Pre  & Sen  & F1   & Spe  & Gm   \\
\hline
0.62 & 0.76 & 0.99 & 0.62 & 0.76 & 0.86 & 0.73 \\
0.53 & 0.33 & 0.96 & 0.53 & 0.69 & 0.27 & 0.38 \\
0.03     & 0.35     & 0.75     & 0.01     & 0.01     & 0.98     & 0.04     \\
0.03     & 0.35     & 1.00     & 0.01     & 0.01     & 1.00     & 0.06     \\
0.03     & 0.37     & 0.90     & 0.01     & 0.01     & 0.98     & 0.07     \\
\hline
\end{tabular}%
}}

\end{table*}
Across all models and datasets, \truevul{} consistently outperforms the baselines, including Naive, PerturbVar, PerturbAPI, and PerturbJoint, demonstrating both strong effectiveness and robustness under different settings.

\subsubsection{\textbf{Intra-project effectiveness}}

As shown in Table~\ref{table:full_expr_results}, \truevul{} achieves the best results across all detectors, with Accuracy of 0.74--0.84, AUC of 0.77--0.86, and F1 of 0.82--0.91. 
Importantly, the effectiveness is achieved with high Precision of 0.96--0.99 while maintaining balanced Sensitivity of 0.71--0.85 and Specificity of 0.72--0.89, resulting in the highest G-mean of 0.78--0.80 among all approaches. 
These results indicate that \truevul{} effectively identifies both untrustworthy and trustworthy predictions.

In contrast, the baselines exhibit unstable and often poor effectiveness. 
Naive achieves high Precision of 0.82--0.94 but suffers from low Sensitivity of 0.31--0.58 and Specificity of 0.32--0.57.
This leads to significantly lower Accuracy of 0.32--0.58, AUC of 0.31--0.60, and F1 of 0.47--0.69.
The perturbation-based methods, PerturbVar, PerturbAPI, and PerturbJoint, frequently achieve high Specificity of 0.70-1.00 at the cost of near-zero Sensitivity. 
They are biased to classify predictions as trustworthy.
Their Accuracy, AUC, F1 and G-mean are consistently low, suggesting that these methods cannot effectively detect untrustworthy predictions.

\subsubsection{\textbf{Cross-project effectiveness}}

Table~\ref{table:full_expr_results_cp} reports cross-project effectiveness on MegaVul, SARD, and PrimeVul.
Despite encountering unseen vulnerabilities, \truevul{} maintains strong effectiveness, achieving Accuracy of 0.62--0.85, AUC of 0.70--0.91, and F1 of 0.76--0.92 across all settings.
\truevul{} still maintains high Precision of 0.98--1.00, Sensitivity of 0.62--0.86, Specificity of 0.61--1.00, and G-mean of 0.68--0.91.
The consistently good effectiveness confirms that \truevul{} is robust and can generalise to unseen vulnerabilities.
By comparison, all baselines exhibit similar trends as in the intra-project setting.

\subsection{Effectiveness Across Vulnerability Types}
\begin{table}[h]
\caption{\truevul's effectiveness on Top 25 Most Dangerous Software Weaknesses in 2025~\cite{CWE:2024:Top25}}
\label{table:per_vul_type}
\centering
\setlength\tabcolsep{3.5pt} 
\begin{tabular}{|l|rrrr|CPP|}
\hline
\multicolumn{1}{|c|}{\textbf{CWE-ID}} & \multicolumn{1}{c}{\textbf{TP}} & \multicolumn{1}{c}{\textbf{TN}} & \multicolumn{1}{c}{\textbf{FP}} & \multicolumn{1}{c|}{\textbf{FN}} & \multicolumn{1}{c}{\textbf{ACC}~$\uparrow$} & \multicolumn{1}{c}{\textbf{FPR}~$\downarrow$} & \multicolumn{1}{c|}{\textbf{FNR}~$\downarrow$} \\
\hline
CWE-20                              & 25716                           & 889                             & 165                             & 6799                            & 0.79                           & 0.16                             & 0.21                             \\
CWE-119                             & 17557                           & 1082                            & 149                             & 4758                            & 0.79                           & 0.12                             & 0.21                             \\
CWE-78                              & 12702                           & 848                             & 139                             & 5107                            & 0.72                           & 0.14                             & 0.29                             \\
CWE-787                             & 10335                           & 780                             & 104                             & 4219                            & 0.72                           & 0.12                             & 0.29                             \\
CWE-125                             & 5478                            & 106                             & 42                              & 1252                            & 0.81                           & 0.28                             & 0.19                             \\
CWE-190                             & 3320                            & 84                              & 29                              & 822                             & 0.80                           & 0.26                             & 0.20                             \\
CWE-400                             & 2728                            & 136                             & 33                              & 805                             & 0.77                           & 0.20                             & 0.23                             \\
CWE-476                             & 1313                            & 70                              & 14                              & 236                             & 0.85                           & 0.17                             & 0.15                             \\
CWE-416                             & 1252                            & 60                              & 28                              & 332                             & 0.78                           & 0.32                             & 0.21                             \\
CWE-200                             & 949                             & 84                              & 10                              & 308                             & 0.76                           & 0.11                             & 0.25                             \\
CWE-77                              & 542                             & 12                              & 10                              & 91                              & 0.85                           & 0.45                             & 0.14                             \\
CWE-22                              & 187                             & 14                              & 2                               & 61                              & 0.76                           & 0.13                             & 0.25                             \\
CWE-287                             & 164                             & 4                               & 5                               & 50                              & 0.75                           & 0.56                             & 0.23                             \\
CWE-79                              & 89                              & 17                              & 6                               & 60                              & 0.62                           & 0.26                             & 0.40                             \\
CWE-269                             & 114                             & 2                               & 0                               & 29                              & 0.80                           & 0.00                             & 0.20                             \\
CWE-89                              & 50                              & 1                               & 0                               & 10                              & 0.84                           & 0.00                             & 0.17                             \\
CWE-863                             & 43                              & 0                               & 0                               & 9                               & 0.83                           & x            & 0.17                             \\
CWE-94                              & 14                              & 7                               & 1                               & 7                               & 0.72                           & 0.13                             & 0.34                             \\
CWE-862                             & 28                              & 0                               & 0                               & 9                               & 0.76                           & x            & 0.24                             \\
CWE-502                             & 12                              & 0                               & 0                               & 5                               & 0.71                           & x            & 0.29                             \\
CWE-434                             & 11                              & 0                               & 0                               & 2                               & 0.85                           & x            & 0.15                             \\
CWE-918                             & 4                               & 0                               & 0                               & 4                               & 0.50                           & x            & 0.50                             \\
CWE-306                             & 4                               & 0                               & 2                               & 2                               & 0.50                           & 1.00                             & 0.33                             \\
CWE-352                             & 3                               & 0                               & 0                               & 0                               & 1.00                           & x            & 0.00                             \\
CWE-798                             & 0                               & 0                               & 0                               & 0                               &           & x            & x           \\
\hline
\end{tabular}
\end{table}

\begin{figure}[!t]
    \centering
    \includegraphics[width=\linewidth, clip, trim=0 0 0 9mm]{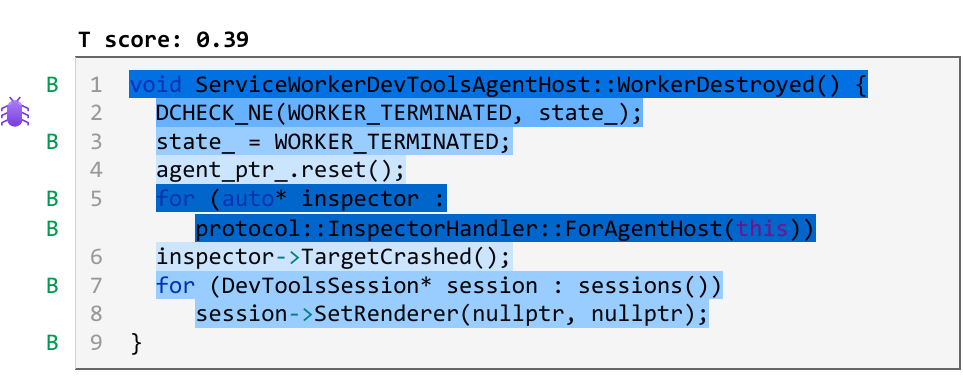}
    
    \vspace{-\baselineskip}
    \caption{TP of IVDetect on SARD with CWE-20.}
    \label{fig:qual_tp}
    \vspace{\baselineskip}

    \centering
    \includegraphics[width=\linewidth, clip, trim=0 0 0 9mm]{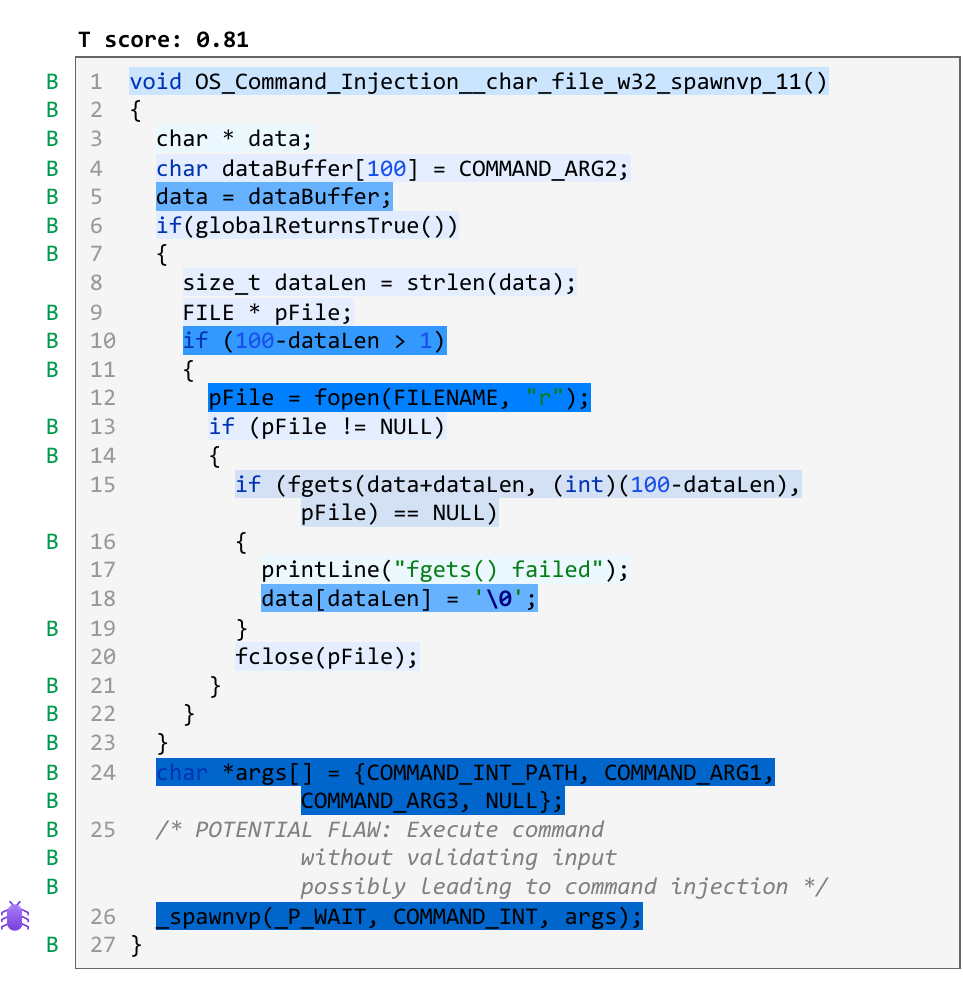}

    \vspace{-\baselineskip}
    \caption{TN of LineVul on BigVul with CWE-78.}
    \label{fig:qual_tn}
    \vspace{\baselineskip}

    \centering
    \includegraphics[width=\linewidth, clip, trim=0 0 0 9mm]{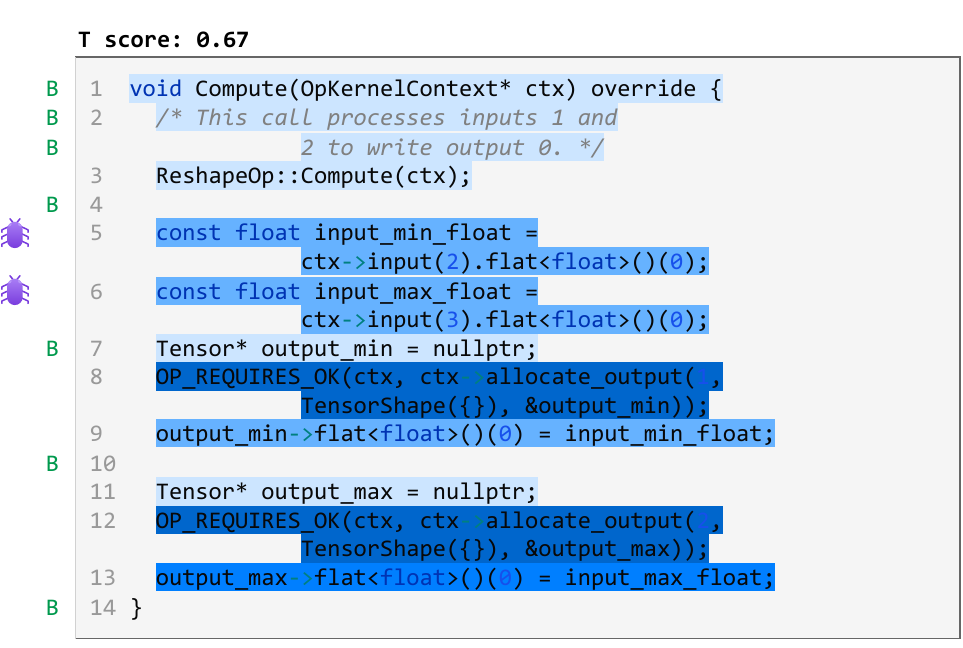}

    \vspace{-\baselineskip}
    \caption{FN of SVulD on BigVul with CWE-787.}
    \label{fig:qual_fn1}
    \vspace{-5mm}
\end{figure}

\begin{figure}[!t]
    \centering
    \includegraphics[width=\linewidth, clip, trim=0 0 0 9mm]{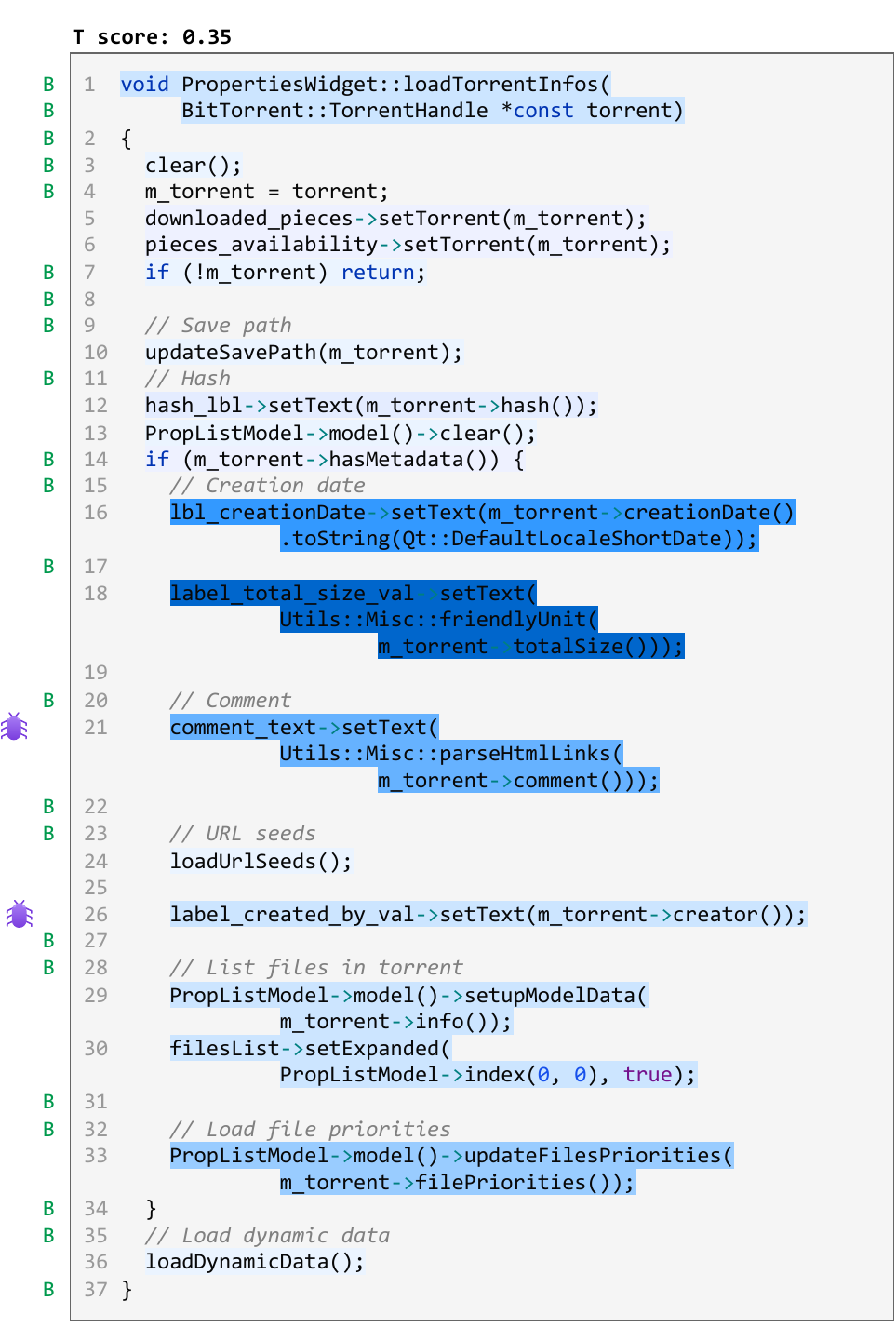}

    \vspace{-\baselineskip}
    \caption{FN of LineVul on MegaVul with CWE-79.}
    \label{fig:qual_fn2}
    \vspace{\baselineskip}

    \centering
    \includegraphics[width=\linewidth, clip, trim=0 0 0 9mm]{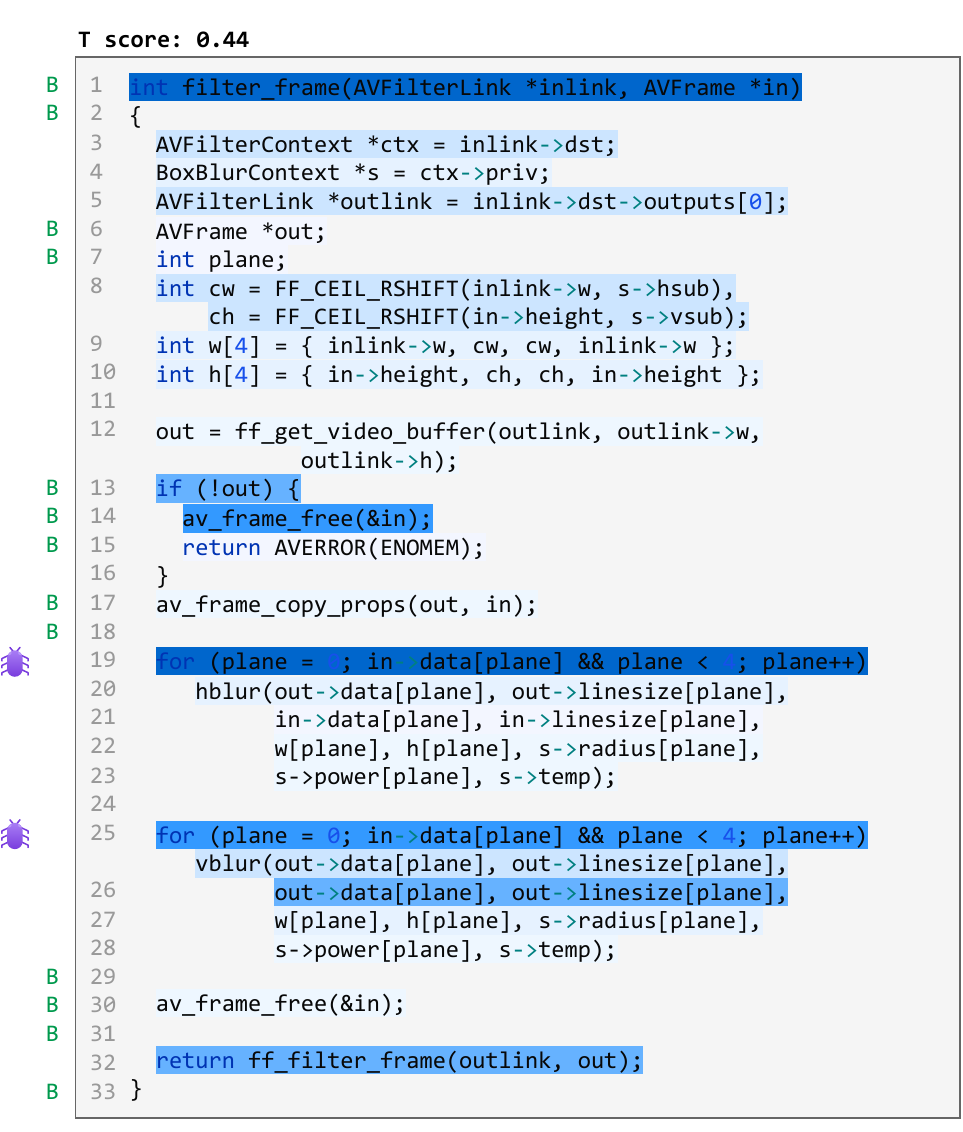}

    \vspace{-\baselineskip}
    \caption{FP of ReVeal on MegaVul with CWE-119.}
    \label{fig:qual_fp}
\end{figure}

Table~\ref{table:per_vul_type} presents \truevul's performance on the top 25 most dangerous types of software vulnerabilities in 2024.
We report metrics including true positives (TP), true negatives (TN), false positives (FP), and false negatives (FN) (i.e., untrustworthy predictions correctly identified, trustworthy predictions correctly identified, trustworthy predictions misclassified as untrustworthy, untrustworthy predictions misclassified as trustworthy, respectively).
We also include overall Accuracy, false positive rate (FPR), and false negative rate (FNR).
Results show that \truevul{} typically achieves Accuracy of 75\%--85\%, FPR of 10\%--32\%, and FNR of 15\%--29\%, except for certain CWE-IDs with only very few samples.

\subsection{Manual Analysis}

We also manually examine several TP, TN, FP, and FN cases, as shown in Figures~\ref{fig:qual_tp}--\ref{fig:qual_fp}.
Lines of code are highlighted based on their importance scores assigned by the models, with darker shading indicating higher importance.
The left column marks whether a line is classified as a benign candidate (denoted by B) by the ensemble model in line-level assessment, and highlights vulnerable lines with a bug icon.

In the TP case shown in Figure~\ref{fig:qual_tp}, the function has an Improper Input Validation vulnerability (CWE-20) at Line~2.
IVDetect correctly detects this function as vulnerable, mostly relying on Lines~1 and~5, which are irrelevant to the vulnerability.
\truevul{} calculates the trustworthiness score $\mathcal{T} = 0.39$, correctly flagging the prediction as untrustworthy.
In the TN case shown in Figure~\ref{fig:qual_tn}, the function contains an OS Command Injection vulnerability (CWE-78) at Line~26.
LineVul correctly identifies this function as vulnerable, paying most attention to Lines~24 and~26.
The computed trustworthiness score $\mathcal{T} = 0.81$, indicating its trustworthiness.

In Figure~\ref{fig:qual_fn1}, the function is vulnerable at Lines~5--6 with the type of Out-of-bounds Write (CWE-787).
The prediction made by SVulD is correct, but based on irrelevant lines like Lines~8,~12, and~13, making it untrustworthy.
\truevul{}, however, fails to flag this prediction as trustworthy.
It assigns a high trustworthiness score $\mathcal{T} = 0.70$, as it does not recognise these irrelevant lines as benign.
Similarly, the function in Figure~\ref{fig:qual_fn2} is vulnerable at Lines~21, and~26 with the type of Cross-site Scripting (CWE-79).
LineVul correctly detect this function as vulnerable, but this prediction is untrustworthy due to overreliance on Lines~16, and~18.
\truevul{} fails to detect these lines as benign candidates.
As a result, it calculates a trustworthiness score $\mathcal{T} = 0.85$, thinking the prediction is trustworthy.
\truevul{} occasionally fails to detect untrustworthy predictions.
We believe the main reason is that it may fail to distinguish between syntactically similar benign and vulnerable lines.

A FP case is illustrated in Figure~\ref{fig:qual_fp}, with Buffer Overflow vulnerabilities at Lines~19 and~25.
Although ReVeal annotates these vulnerable lines, it relies on many irrelevant lines classified as benign candidates by \truevul, such as Lines~1,~13,~14,~32.
Their impact outweighs the vulnerable lines.
Hence, this prediction is deemed untrustworthy with the trustworthiness score $\mathcal{T} = 0.44$.

\begin{figure}[ht]
    \centering
    \includegraphics[width=\linewidth, clip, trim=0 0 0 9mm]{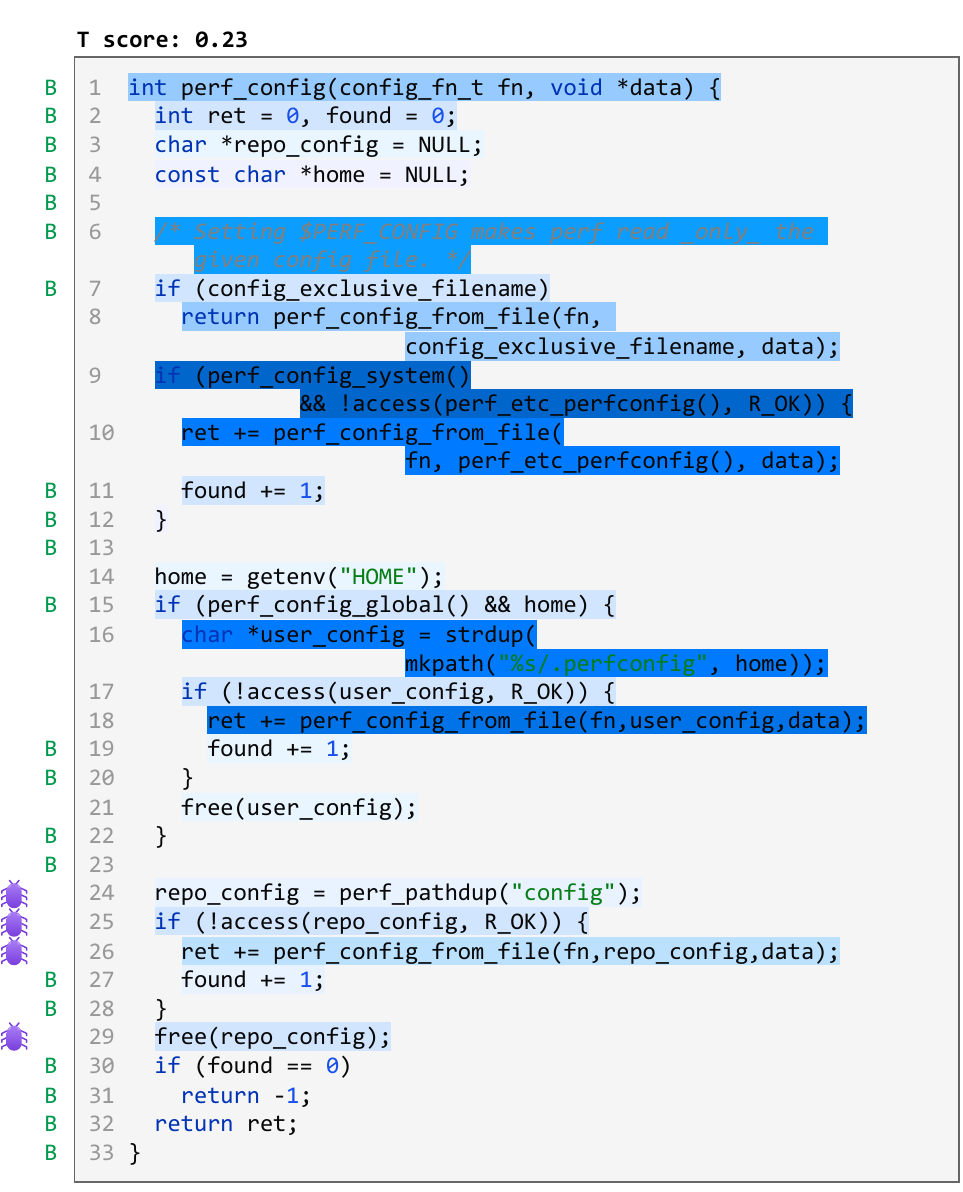}

    \vspace{-\baselineskip}
    \caption{An edge case of untrustworthiness where the suspicious lines highlighted by LineVul are not actual vulnerable lines but syntactically similar to them.}
    \label{fig:edgecase}
\end{figure}

We also analyse untrustworthy cases detected by \truevul{} but missed by the baselines. 
Compared with the baselines, \truevul{} detects a total of 40,251 additional untrustworthy predictions. 
Among these, we identify 816 edge cases, where dependencies are ambiguous or suspicious lines of code highlighted by the detectors are not actual vulnerable lines but merely syntactically similar to vulnerable ones.
Figure~\ref{fig:edgecase} illustrates one such edge case, where the prediction is made by LineVul for the function \textcode{perf\_config}. 

The function \textcode{perf\_config} looks for configuration files in several locations: \textcode{/etc/perfconfig} at Lines~7--12, \textcode{\$HOME/.perfconfig} at Lines~14--22, and \textcode{config} in the current working directory at Lines~24--29.
Unlike the first two paths, ``config'' is not an unusual name for a file that may already exist.
Hence, if \textcode{./config} is not a valid perf configuration file, the program would fail, or worse, mistakenly treat it as a perf configuration file and change the program's behavior in some unexpected way.
Consequently, the vulnerability occurs at Lines~24--29.
Although LineVul correctly identifies this function as vulnerable, it mainly highlights Lines~9,~18,~10,~16, and~6 (listed in descending order of importance scores).
The highlighted suspicious lines are clustered into two regions where configuration is loaded from the files \textcode{/etc/perfconfig} and \textcode{\$HOME/.perfconfig}.
These lines are unrelated to the actual vulnerability, nor do they all share dependencies with one another.
As a result, the prediction is \textit{untrustworthy}.

Notably, in this untrustworthy prediction, Lines~18 and~10 are syntactically similar to Line~26, and Line~9 is partially similar to Line~25.
If this untrustworthy prediction goes undetected, developers can be misled into believing that the vulnerability is associated with the configuration in \textcode{/etc/perfconfig} or \textcode{\$HOME/.perfconfig}.
The baselines fail to detect this prediction as untrustworthy, but \truevul{} correctly identifies it as such.
As a result, \truevul{} would have saved developers from chasing irrelevant lines of code, especially in the edge cases like the one in Figure~\ref{fig:edgecase}, which are difficult to detect and can easily mislead developers.

\end{document}